\documentclass[]{aa}
\usepackage{graphicx}
\usepackage[varg]{txfonts}
\usepackage{natbib}
\bibpunct{(}{)}{;}{a}{}{,} 
\usepackage[breaklinks]{hyperref}
\usepackage{breakurl}
\def\teff{$T\rm_{eff}$}

\newcommand{\glog}{\ensuremath{\log {\rm g}}}
\newcommand{\cobold}{{\sf CO$^5$BOLD}}
\renewcommand{\lhd}{{\sf LHD}}
\newcommand{\linfor}{Linfor3D}

\idline{100}{100}
\begin{document}

\title{Using the CIFIST grid of CO5BOLD 3D model atmospheres to study the effects of stellar granulation on photometric colours.  }
\subtitle{I. Grids of 3D corrections in the UBVRI, 2MASS, Hipparcos, Gaia, and SDSS systems.}

\author{
P. \,Bonifacio\inst{1} \and
E.\,Caffau\inst{1} \and
H.-G.~Ludwig  \inst{2,1} \and
M.\,Steffen\inst{3,1}\and
F.\,Castelli\inst{4}\and
A.\,J.\,Gallagher \inst{5,1}\and
A.\,Ku\v{c}inskas\inst{6}\and
D.\,Prakapavi\v{c}ius\inst{6}\and
R.\,Cayrel\inst{1} \and
B.\,Freytag\inst{7}\and
B.\,Plez\inst{8}\and
D.\,Homeier\inst{2}
}
\institute{ 
GEPI, Observatoire de Paris,  PSL Research University, CNRS,  
Place Jules Janssen, 92195 Meudon, France
\and
Zentrum f\"ur Astronomie der Universit\"at Heidelberg, Landessternwarte, 
K\"onigstuhl 12, 69117 Heidelberg, Germany
\and
Leibniz-Institut f\"ur Astrophysik Potsdam, An der Sternwarte 16, D-14482 Potsdam, Germany
\and
Istituto Nazionale di Astrofisica, 
Osservatorio Astronomico di Trieste, Via Tiepolo 11,
I-34131 Trieste, Italy
\and
Max Planck Institut f\"ur Astronomie, D-69117 Heidelberg, Germany
\and
Astronomical Observatory, Institute of Theoretical Physics and Astronomy, Vilnius University, Saul\.{e}tekio al. 3, Vilnius LT-10257, Lithuania
\and
Department of Physics and Astronomy at Uppsala University, Regementsv{\"a}gen 1, Box 516, 75120, Uppsala, Sweden
\and
LUPM,  Universit\'e de Montpellier, CNRS, UMR 5299, 34095 Montpellier Cedex 05, France
}
\authorrunning{Bonifacio et al.}
\titlerunning{colours}
\offprints{P.~Bonifacio}
\date{Received ...; Accepted ...}

\abstract%
{The atmospheres of cool stars 
are temporally and spatially inhomogeneous
due to the effects of convection.  
The influence of this inhomogeneity, referred
to as granulation, on colours  has never 
been investigated over a large range of effective temperatures and gravities.
}
{We aim to study, in a quantitative way, the impact
of granulation on colours.
}
{
We use the CIFIST (Cosmological Impact of the FIrst Stars) grid of 
\cobold\ (COnservative COde for the COmputation of COmpressible COnvection in a BOx of L Dimensions, L=2,3)
hydrodynamical models to compute emerging fluxes. 
These in turn are used to compute theoretical colours in the $UBVRI$, 2MASS, Hipparcos, Gaia 
and SDSS systems. 
Every \cobold\ model has a corresponding one dimensional (1D) plane-parallel 
\lhd\ (Lagrangian HydroDynamics) model computed for the same atmospheric parameters, 
which we used to  define a "3D correction" that can
be applied to colours computed from fluxes
computed from any 1D model
atmosphere code. As an example, we illustrate
these corrections applied to colours computed from ATLAS models.
}
{
The 3D corrections on colours are generally small, of the order of a few hundredths of a magnitude, yet they are far from negligible. We find that ignoring granulation effects can lead to underestimation of \teff\ by up to 200\,K and overestimation of gravity by up to 0.5\,dex, when using colours
as diagnostics. 
We have identified a major shortcoming in how scattering is 
treated in the current version of the CIFIST grid, which could lead to offsets of the order 0.01\,mag, especially for colours involving blue and UV bands. 
We have investigated the Gaia and Hipparcos photometric systems
and found that the  $(G-H_p)$,$(BP-RP)$ diagram
is  immune to the effects of granulation. In addition, we point to  
the potential of the RVS photometry as a metallicity diagnostic.
}
{Our investigation shows that the effects
of granulation should not be neglected if one
wants to use colours as diagnostics of the stellar
parameters of F,G,K stars. A limitation is that scattering is treated as true absorption
in our current computations, thus our 3D corrections are likely
an upper limit to the true effect.
We are already computing the next generation
of the CIFIST grid, using an approximate treatment
of scattering. }
\maketitle

\section{Introduction}

The use of theoretical colours computed from model atmospheres
is widespread, however, so far all available
grids of theoretical colours \citep[e.g.][]{Bessel98,Castelli99,onehag,Casagrande14}
are based on one dimensional (1D) static model atmospheres. 
Extensive computations of fluxes from the three-dimensional (3D) models of the 
{\tt STAGGER} grid \citep{Magic1} have been done by
\citet{Magic4}, however, so-far only the limb-darkening coefficients
have been published.
In a series of two  papers, we wish to explore
the parameter space covered by  
the CIFIST (Cosmological Impact of the FIrst Stars)  
grid of 3D  models \citep[][and Ludwig et al., in preparation]{cifist09} 
that has been computed with the \cobold\ (COnservative COde for the COmputation of COmpressible COnvection in a BOx of L Dimensions, L=2,3) 
code \citep{FSL12}.
This effort expands on the pioneering computations of colours 
from \cobold\ models  presented earlier 
by \citet{KHL05} and \citet{KLC09}.

For comparison with the \cobold\ models, 
we computed two grids of 1D model atmospheres. 
The first was computed using the \lhd\  (Lagrangian HydroDynamics) 1D model atmosphere code 
\citep{CL07} assuming a mixing length parameter $\rm\alpha_{LHD} = 1.0$ 
\footnote{Associated to  each model in the CIFIST grid we always compute
three \lhd\ models with  $\rm\alpha_{LHD} = 0.5, 1.0, 2.0$.  }. 
 The value 1.0
is closest to the value used in the \citet{CK03} grid of ATLAS models; therefore
we chose this value here. The effects of different choices
of the mixing length parameter  in the reference  1D model are further explored in Paper II of the series \citep{paper2}. 
The second grid was computed using ATLAS \citep{K05} assuming a mixing length parameter 1.25,
and the \citet{CK03} Opacity Distribution Functions (ODFs)
with 2\,kms$^{-1}$ microturbulence.  The choice of mixing
length and microturbelent velocity were made to be consistent
with the  \citet{CK03} grid of ATLAS models.
Both grids of models were specifically 
computed to match atmospheric parameters of the \cobold\ models 
in the CIFIST grid.

The purpose of the present  paper (Paper I of the series) 
is to present the computational methods employed and compare the fluxes and
colours computed from 1D static model atmospheres to those computed
from 3D hydrodynamical model atmospheres.
We discuss in detail the effects of the treatment of scattering both
in the computation of model atmospheres and in the computation of emerging
fluxes. 
For the $UBVRI$, 2MASS, Hipparcos, Gaia and SDSS $ugriz$
systems we provide grids of ``3D corrections'' that can be applied to 
colours computed from 1D model atmospheres. We also provide a corresponding
grid of ATLAS 3D-corrected colours.

In the second paper of the series \citep{paper2}, 
we select ten model atmospheres
for two different metallicities (0.0 and --2.0) that cover the main phases
of stellar evolution (main sequence, turn off, sub giant, red giant) and explore
the effects of granulation in the corresponding parameter space.

There are  some differences of approach to the computation of colours in the
two papers, which are explained in detail in \citet{paper2}.  
In both papers, we adopt a strictly differential approach, discussing only 
3D-1D differences, which makes the above-mentioned differences irrelevant.

\section{Computation of synthetic colours}

\subsection{Photometric systems and zero-points}

\begin{table}
\centering
\caption{2MASS in-band fluxes and corresponding magnitudes from \citet{Cohen}\label{2mass}}
\renewcommand{\tabcolsep}{3pt}
\begin{tabular}{ccccccc}
\hline
Star   & $FJ$ & $J$     & $FH$  & $H$  &   $FK_S$ & $K_S$  \\
       & $\rm Wcm^{-2}$  & mag & $\rm Wcm^{-2}$ & mag &$\rm Wcm^{-2}$ & mag \\
       &$\times 10^{-14}$& &$\times 10^{-14}$& &$\times 10^{-14}$& \\  
\hline\hline
Vega   & \phantom{1}5.076 & \phantom{--}0.001& \phantom{1}2.857 & --0.005 & 1.121 & \phantom{--}0.001 \\
0 mag  & \phantom{1}5.082 & \phantom{--}0.000 & \phantom{1}2.843 & \phantom{--}0.000 & 1.122 & \phantom{--}0.000 \\
\hline\hline
\end{tabular}
\end{table}

The problem of computing synthetic colours
has been thoroughly addressed 
in many excellent papers
\citep[e.g.][]{Bessel98,Castelli99,onehag,CK06,BM12,Casagrande14}.
Our own approach to the problem has been extensively described
in \citet{Bonifacio_Mem}.
We refer the reader to that paper for justification on our choices of 
zero points for the synthetic photometry and bolometric magnitudes.
In this paper, we investigate the $UBVRI$ and Hipparcos-Tycho systems, 
using the bandpasses defined by \citet{BM12}, the 2MASS system
\citep{Cohen}  and the Gaia system, as defined by the pre-launch
bandpasses \footnote{\url{http://www.cosmos.esa.int/documents/29201/302420/normalisedPassbands.txt/a65b04bd-4060-44fa-be36-91975f2bd58a}}.
The above are all ``Vega'' systems, and we tie the synthetic photometry
to Vega using the CALSPEC flux of VEGA\footnote{\url{ftp://ftp.stsci.edu/cdbs/current_calspec/alpha_lyr_stis_008.fits}}
and assuming all magnitudes for Vega to be 0.03, except for the 2MASS
magnitudes where we assume for Vega J=0.001, H=--0.005 and K=0.001,
in order to be consistent with the zero-magnitude fluxes of \citet[][see table \ref{2mass}]{Cohen}. 
As explained in  \citet{Bonifacio_Mem}, we did not define the zero points
of the ``Vega'' systems by using a model atmosphere of Vega,
since such a model cannot be computed with \cobold , but we
assumed a radius of $1\,R_\odot$ and a distance of 10\,pc
for each model atmosphere and applied to this flux the same
zero points adopted for the observed fluxes. 
We also investigated the SDSS $ugriz$ system, which  is  AB type 
\citep{OkeGunn83}, tied to the standards defined in \citet{Fukugita}.
In the SDSS catalogue, what are reported are  not magnitudes, 
but  ``luptitudes'' defined in \citet{Lupton}.
A ``luptitude'' is defined as
\begin{equation}
m = -{2.5\over\ln(10)}\left[{\rm asinh}\left((f/f_0)/(2b)\right)+\ln(b)\right]
\label{luptitude}
\end{equation}
where $f$ is the flux of the object ,$f_0$ is the flux
of the 0 magnitude object, and $b$ is a constant, called the softening
parameter. 
The softening parameter determines the magnitude 
of an object with zero flux, that is, a non-detection in the survey; this is
 the magnitude limit of the survey. 
It is important to keep in mind that objects with magnitudes
equal or very close to this limit have essentially undefined
colours and therefore should not be compared to theoretical colours.
Other SDSS-like systems that are becoming increasingly
popular and are used in many telescopes may use magnitudes rather than luptitudes.
In order to be able to compare our computed colours directly to the SDSS catalogue, 
we computed luptitudes rather than magnitudes to derive SDSS  colours. 
We note however that the difference in 3D correction between colours computed
using luptitudes and magnitudes is always less than or equal to $10^{-5}$ mag.
So for any practical purpose the 3D corrections provided in Tables \ref{SDSS_1_m00}
to \ref{SDSS_2_m30} can also be applied to any SDSS-like system.

\subsection{Computation of emerging fluxes}

As described in greater detail in \citet{Bonifacio_Mem}
we used the {\tt NLTE3D} code to compute the emerging
fluxes from the \cobold\ models. 
Each 3D model is formed by a time series of 3D structures, usually of the order of several hundreds, that we call snapshots.
Consecutive snapshots are often correlated, and including correlated events does not improve the  statistic.
As already mentioned, computations with 3D structures, whether this be
fluxes or spectrum synthesis,  are time consuming.
We then  select a representative number of statistically independent snapshots able to represent the model.
Usually we select about 20 snapshots from a model, because we think this is a good compromise between the computation time for the fluxes or spectrum synthesis, 
and the representativeness of the selection for the model. 
This number has been found to be representative in a detailed
investigation on the solar model
\citep{tesidetta} and we assume this to be the case for any \cobold\ model.
The snapshot selection is made so that its statistical
properties are the same as those of the total ensemble
of computed snapshots; we refer to \citet{CL07} for 
further details on this. 
For each model, 
we averaged the emerging flux from the selected snapshots.

We relied
on the \citet{CK03} opacity distribution functions (ODFs) in 
order to take into account the line opacity. 
We used the ``LITTLE'' 
ODFs, with 1212 frequency points, rather than the ``BIG'' 
ODFs that only have 328 frequency points.
The continuum opacities 
were computed using the {\tt iondis} and {\tt opalam} 
routines of the \linfor\ spectrum synthesis code \citep{Steffen2017,Gal17}.

This approach is inconsistent, since the models have been
computed using MARCS continuum and line opacities,
while the fluxes have been computed assuming  {\tt iondis} and {\tt opalam}
continuum opacities and ATLAS line opacities. Since
we rely mainly on a differential approach (3D-1D),  we do 
not expect this inconsistency to be a major shortcoming, as we explain below.

\subsection{The treatment of scattering}

Scattering opacity has been treated as true absorption and this, 
in our view is the main physical  limitation of our computation.
The treatment of scattering enters on two occasions in our problem:
first in the computation of the model, where its treatment may influence
the temperature structure of the model, then in the computation of 
the emerging flux from the model.
The choice of treating scattering as true absorption is made
for computational reasons. A fully consistent treatment of scattering
requires that the mean intensity $J_\nu$  be evaluated at each point
in the computational box during the computation.
\citet{collet11} questioned the soundness of  treating
scattering as true absorption
and showed that it may lead to a mean temperature 
structure that is substantially
warmer than that obtained when explicitly treating the scattering. 
On the other hand, they also noticed that a better and computationally less
expensive approximation is to treat scattering as true absorption in the
optically thick layers and ignore the scattering opacity in the 
optically thin layers as described in \citet{Hayek}. 
In this approximation one does not really treat scattering
while computing the radiative transfer, but uses an appropriate 
opacity table.
\citet{LS12} computed a \cobold\ model using the \citet{Hayek} 
approximation
and agreed qualitatively with the conclusions of \citet{collet11},
although in their case the influence of the treatment of scattering
was clearly less important than in the example studied
by \citet{collet11}.  In the  
\citet{LS12} study, the difference in mean temperature between the ``true absorption''
and ``approximate scattering''
models is about 100\,K at log($\tau$)=--4, while in the
\citet{collet11} study, this difference is about five times larger.
The two models used by \citet{LS12} and  \citet{collet11}  
have very similar atmospheric
parameters, therefore we attribute this difference between the two studies
to differences between the two codes used to compute the hydrodynamical 
simulations. 
We have used this model, 
with \teff = 5000 K, log g = 2.5, and metallicity --3.0, and
another model, with the same atmospheric parameters, but computed
treating scattering as true absorption, to assess the influence
of scattering on the computed fluxes and colours.
This particular model was chosen because it was in this
case that \citet{collet11} obtained the largest effects of scattering.
Therefore these results may be considered an upper limit to
the effects of scattering.

\begin{figure}
\centering
\resizebox{\hsize}{!}{\includegraphics[clip=true]{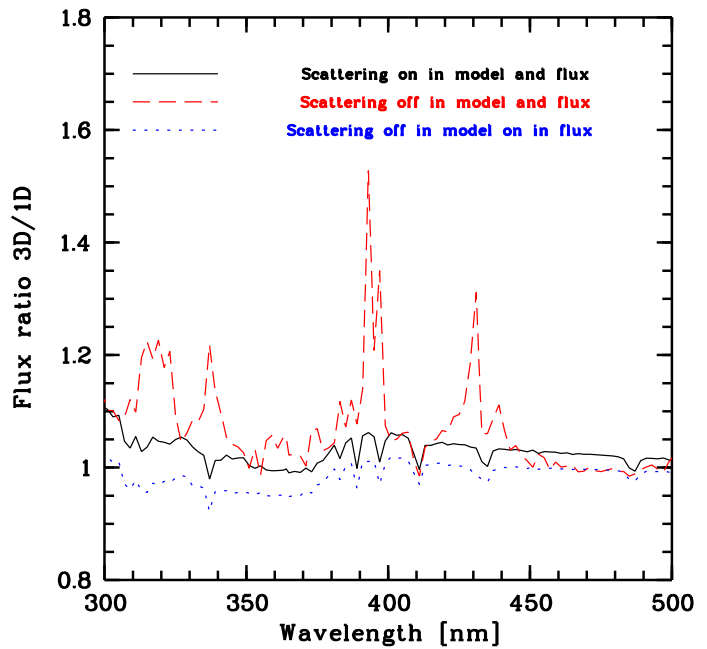}}
\caption{ Ratio of the fluxes of one snapshot of the \cobold\
models d3t50g25mm30n01 (scattering treated as true absorption) and 
d3t50g25mm30n02 (scattering in the \citet{Hayek} approximation) 
to the fluxes computed from
their corresponding \lhd\ models under different assumptions.
The solid line (black) corresponds to the d3t50g25mm30n02 model, 
and the flux has been computed fully taking into account scattering.
The dashed line (red) corresponds to the 
d3t50g25mm30n01 model and the flux has been computed treating scattering
as true absorption. The dotted line (blue) corresponds again to the
d3t50g25mm30n01 model, but in this case scattering has been properly taken into
account in the flux computation.
\label{scattering}}
\end{figure}

{\tt NLTE3D} can treat scattering in detail (no approximations)
in the computation of the flux from a given model, 
but the computation is much more intensive. 
We limited the comparison to a single snapshot. 
In order to check that a single snapshot 
is representative, we computed the colours  for each of the 20 snapshots that are
our selection for this model. The snapshot-to-snapshot difference
in colours was always below 0.01 mag. As we shall see below, the differences
due to the different assumptions on scattering are always larger
than this. Thus our comparison is meaningful. 

In Fig.\,\ref{scattering}, we show the ratio of the flux computed from the \cobold\
models to that of their corresponding \lhd\ models.
The dotted line corresponds to the model for which scattering
has been treated as true absorption, but scattering has been taken into account
in the flux computation. The dashed line corresponds to the same model; 
however in this case, consistently with what was done for the model
computation, also in the flux computation scattering has been treated as true
absorption. Finally the solid line corresponds to the model 
for which  scattering has been
treated in the approximation of \citet{Hayek} and scattering
has been fully taken into account in the flux computation.
The same approximations apply  to both the 1D and 3D models.
We present this in  the UV-visible region where
the differences due to the different treatment of scattering are largest.
The first thing that is obvious is that the treatment of scattering mostly affects 
the spectral regions characterised by strong lines. Treating scattering as
true absorption results in a flux from the \cobold\
model that can be 1.2 to almost 1.6 times larger
than that in the corresponding \lhd\ model. 
This is not so in the case where scattering is treated using the
\citet{Hayek} approximation in the model, and fully treated in the
flux computation. In this case, the flux from the \cobold\ model
is only about 5\% larger than that from the \lhd\ model, but the flux ratio
varies rather smoothly with wavelength, unlike what is seen when 
treating scattering as true absorption.
The ``hybrid'' case, in which we treat scattering in the flux computation using
a model which has been computed treating scattering as true absorption,
is interesting. In the current version of the CIFIST grid \citep{cifist09}
all models have been computed treating scattering as true absorption. 
A new version of the grid is currently being computed in which 
scattering is treated in the \citet{Hayek} approximation.
In the following, for simplicity, 
we refer to the case where
scattering has been treated as true absorption both in the model 
and in the flux calculations as “true absorption”;
we refer to the case in which
the model has been computed treating scattering in the 
\citet{Hayek} approximation and the flux has been computed
fully treating scattering as ``scattering''; we refer to the
case in which the model has been computed treating scattering
as true absorption but the flux has been computed fully taking into
account scattering, as ``hybrid''.  
 
Is there anything to be gained in treating scattering
in the flux computation, from the current version of the CIFIST grid?
Figure\,\ref{scattering} suggests that this is not the case.
Although in this hybrid case the run of the 3D/1D ratio has a shape that
is very similar to that of the scattering case, it is typically
10\% smaller than what is expected in the scattering case.
We are ultimately interested in colours and therefore we computed
the colour-corrections (3D-1D) for all three cases
shown in Fig.\,\ref{scattering}.
Assuming that the scattering computations are the ones
closest to the truth we find 
the difference in 3D correction between
the  scattering  case  
and the true absorption case
is on average --0.013\,mag  while this difference is
 --0.009\,mag for the hybrid case. 
Obviously this is largest, in absolute value,  for the 
bluest colours. For $U-B$ we find a difference in the 3D correction
of --0.024 for the true absorption case and --0.18 for the hybrid
case, while for the $V-K$ colour we find --0.018 for the
true absorption case and --0.014  for the hybrid case. In the wavelength range 500\,nm to 2500\,nm (not shown in Fig. 1) 
the ratio of the scattering 3D/1D flux ratio to the true absorption 
3D/1D flux ratio varies from
0.95 to 1.017
with a mean value of 1.003.

From this test, on one model we decided to use the
fluxes computed from the CIFIST grid treating scattering
as true absorption. 
Although the hybrid approach does present
a flux that has a similar shape to the  consistent scattering
approach, we cannot convincingly conclude that this
improvement is worth the considerable computational effort needed to
perform hybrid computations of the whole grid, considering
that the differences in 3D corrections are of only a few thousandths
of a magnitude.

All fluxes from the ATLAS models were computed with the ATLAS code,
where true absorption was assumed for
the continuum scattering and no line scattering was 
considered. These were also the assumptions adopted in the computation
of the ATLAS models.

\begin{figure}
\centering
\resizebox{\hsize}{!}{\includegraphics[clip=true]{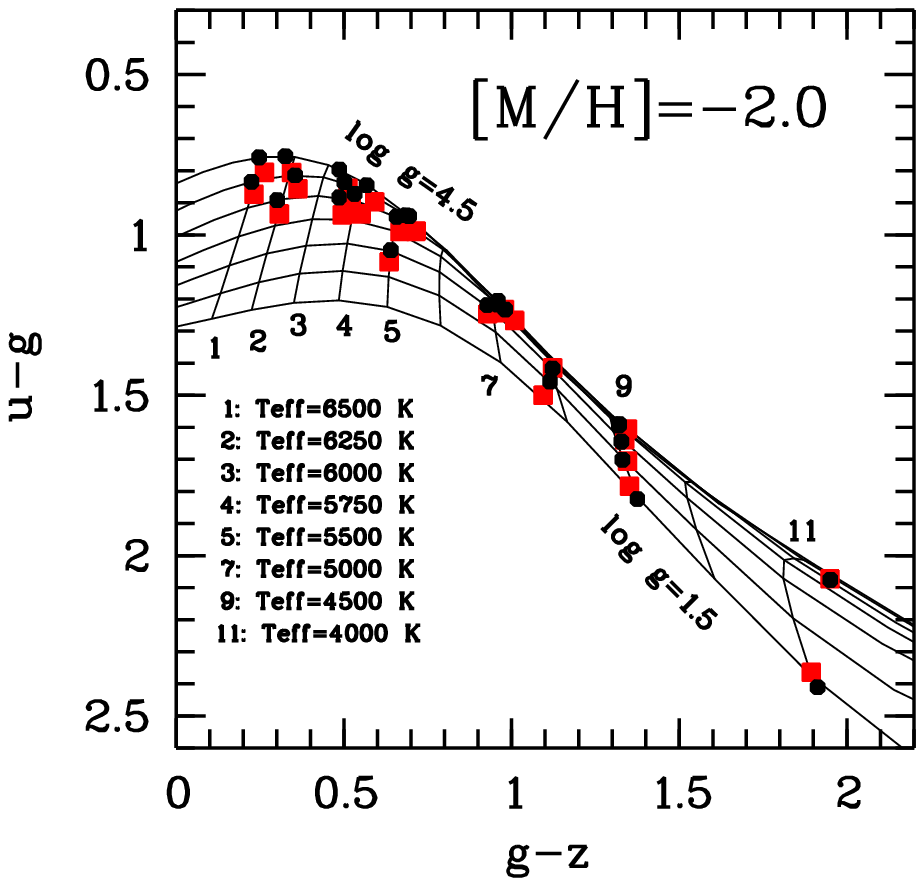}}
\resizebox{\hsize}{!}{\includegraphics[clip=true]{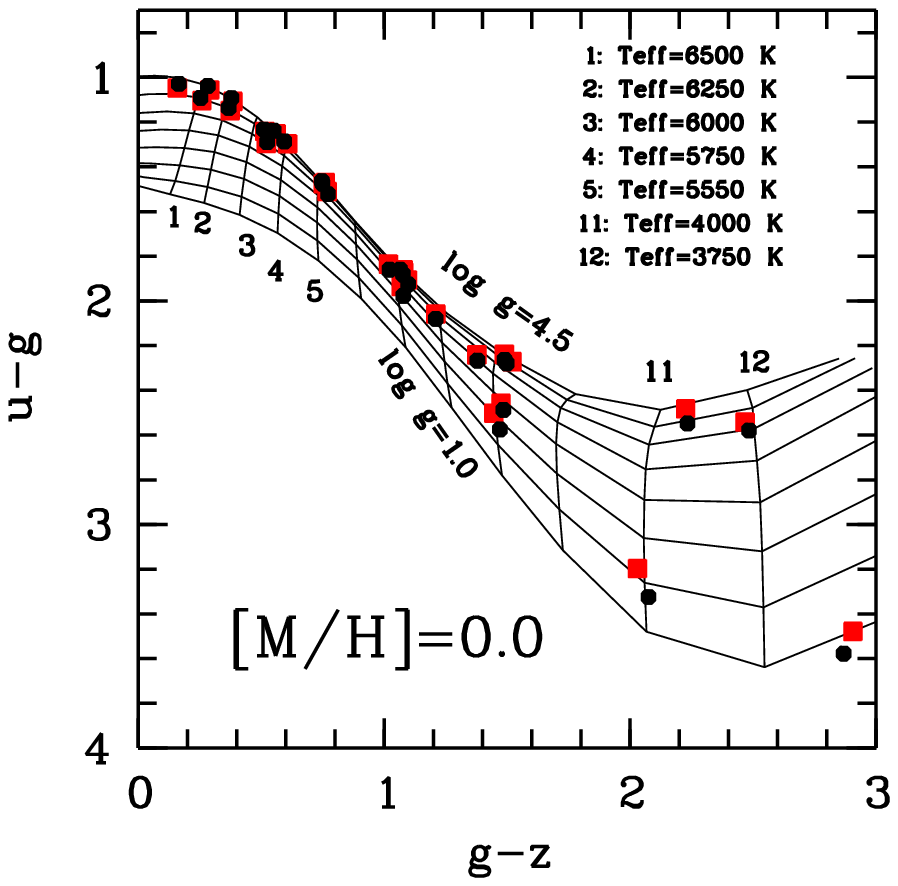}}
\caption{A  $u-g$  vs.$g-z$ colour-colour diagram
for solar metallicity (bottom panel) and --2.0 (upper panel). 
The solid lines are the lines of constant temperature and
constant surface gravity defined by the \citet{CK03}
grid. The black dots correspond to our ATLAS models, that 
sample points within the \citet{CK03} grid that correspond
to the parameters of the models in the CIFIST grid. 
The \citet{CK03} grid has steps of 250\,K in \teff\ and 0.5\,dex
in \glog.
The red squares are the ``3D-corrected'' colours.
\label{ug_gz}}

\end{figure}

\begin{figure*}
\centering
\resizebox{16cm}{!}{\includegraphics[clip=true]{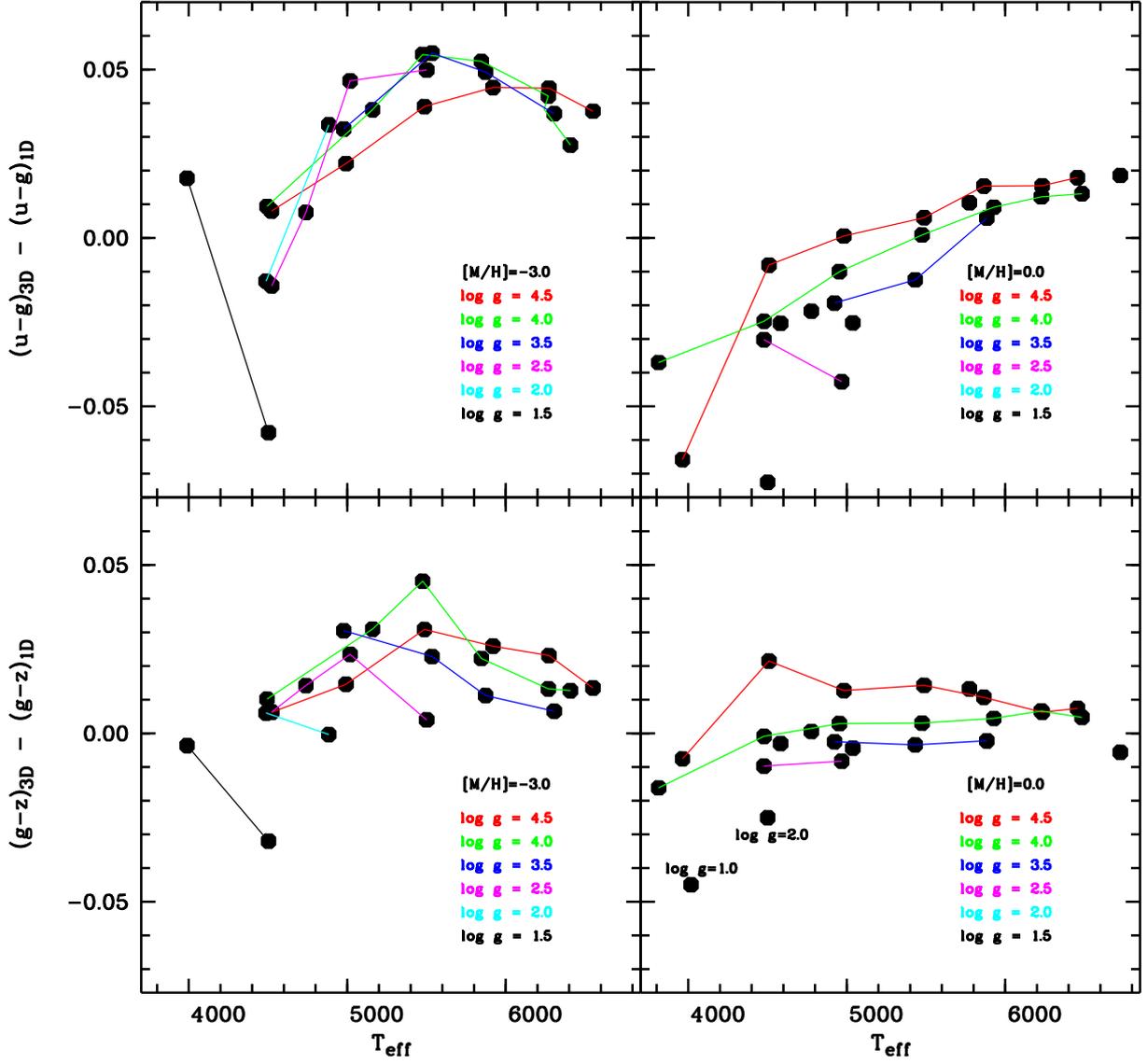}}
\caption{
The 3D--1D  corrections for  the $g-z$ and $u-g$ colours for
two metallicities: 0.0 and --3.0.
\label{coldiff}}

\end{figure*}

\section{Results}

We adopted a differential approach: for each colour, magnitude, or
bolometric correction, we computed the ``3D correction'' as
colour (\cobold ) - colour (\lhd ). These corrections
are provided in the Tables \ref{ubvrim00} to  \ref{SDSS_2_m30}. 
As an example of how these corrections have to be used, we also provide
the ATLAS  3D--corrected colours. 
For each \cobold\ model we computed an ATLAS model
with exactly the same parameters and its emerging flux. 
From this flux, we computed colours and  the 3D corrected colour
is then simply colour (ATLAS) + ``3D correction''.
Clearly, this same procedure can be applied to colours
computed from any grid of 1D model atmospheres.

The ATLAS models computed for this paper are almost identical
to those of the \citet{CK03} grid. The only difference is that 
we used updated ODFs for which the
contribution of the $\rm H_2 O$ lines has been revised. 
This has only a minor effect on models cooler
than 4500\,K. 
 Our computed colours from our ATLAS models 
precisely fall on the theoretical curves defined in the \citet{CK03} 
grid, and may be slightly off for some colours when the temperature decreases
due to H$_2$O opacity.

Figure\,\ref{ug_gz} 
depicts colour-colour diagrams for $g-z$ and $u-g$ 
bands for metallicities at [M/H]=0.0 and --2.0.
If the metallicity is known (e.g. for stars in a cluster)
such a diagram can be used as a diagnostic in \teff\ and \glog .
While the corrections are generally small around the solar 
chemical composition, they increase with decreasing metallicity.
At low metallicity, 
the 3D-corrected $g-z$ tends to be larger (i.e. redder),
therefore, hotter temperatures will be found from observed g-z colours.
At the hot end of our model grid, this change
is rather large, of the order of 200\,K.
At solar metallicity the situation is similar for the hotter
models, however for the cooler models (\teff $<4500$\,K)
the situation reverses and the 3D-corrected $g-z$ is smaller
(bluer); thus, for a given observed $g-z,$ one would infer a cooler
\teff , although by only about 50\,K.
A similar behaviour is observed for the $u-g$ colour at low metallicity; the
3D-corrected $u-g$ is larger, thus, for an observed
$u-g,$ one infers a lower gravity, by almost 0.5\, dex. 
At solar metallicity the situation is different. For the hotter models, there
is hardly any change in $u-g$, but below 5500\,K the 3D-corrected
$u-g$ is smaller, implying gravities that are larger, by about 0.1\,dex.
The situation is illustrated in Fig.\ref{coldiff}, where the 
3D-1D corrections for metallicities 0.0 and --3.0 are plotted as a function
of \teff . To guide the eye, the corrections relative to models 
with the same surface gravity are connected by a solid line and a different
colour is used for each surface gravity as given in the legend.

\begin{figure}
\centering
\resizebox{\hsize}{!}{\includegraphics[clip=true]{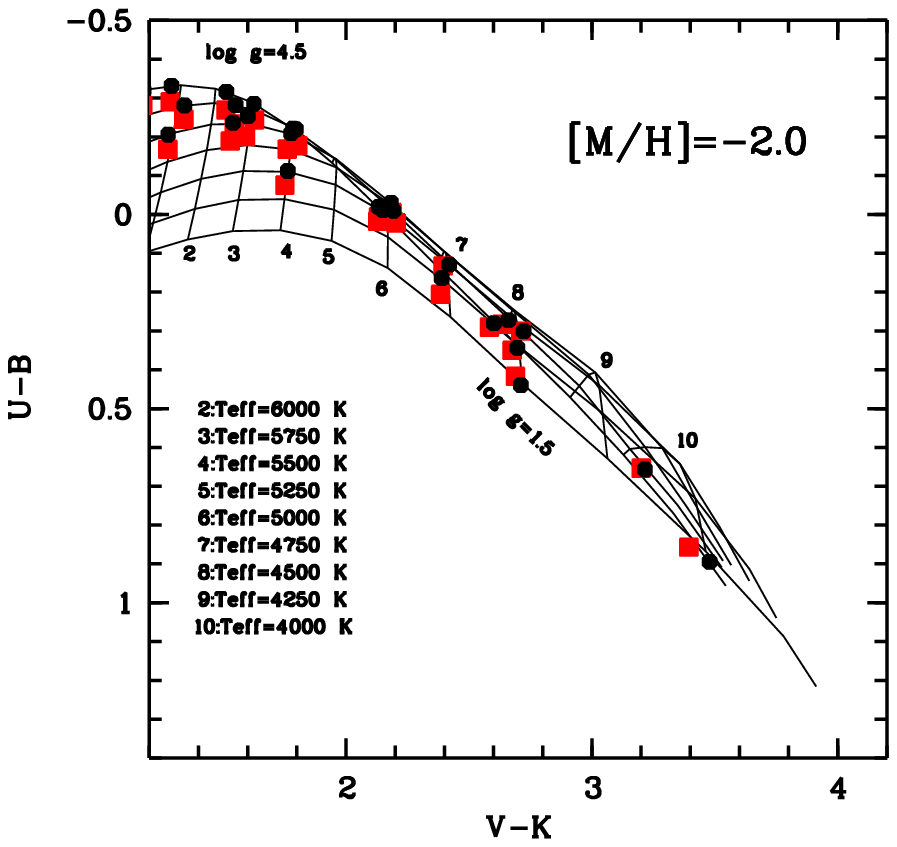}}
\resizebox{\hsize}{!}{\includegraphics[clip=true]{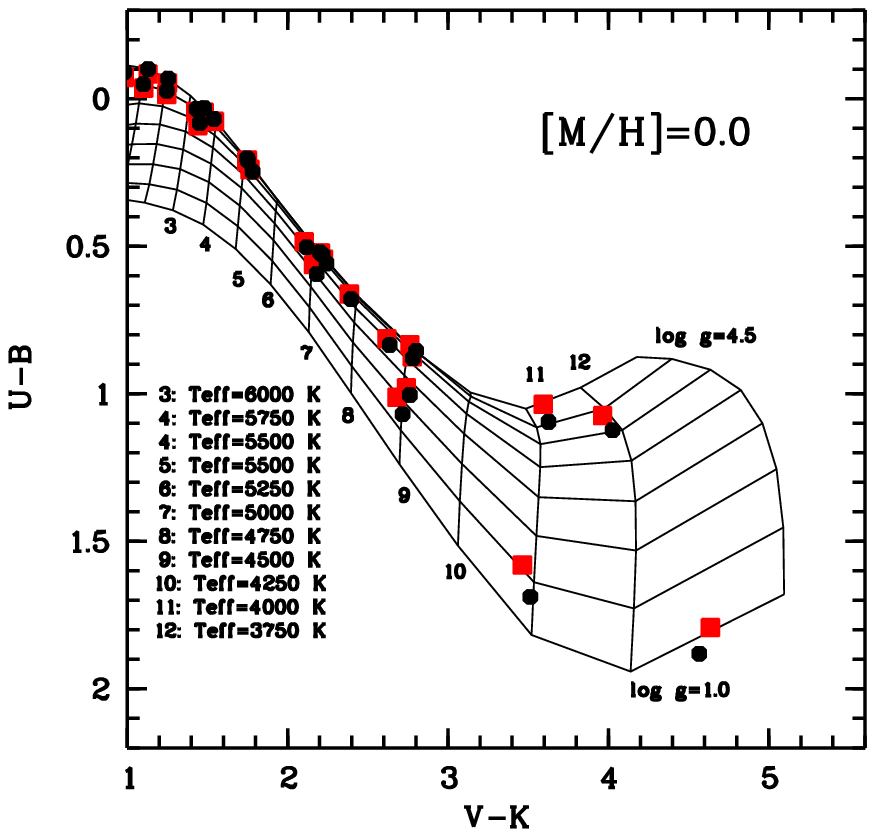}}
\caption{A $U-B$  vs. $V-K$ colour-colour diagram
for solar metallicity (bottom panel) and --2.0 (upper panel). 
The meaning of the lines and symbols is  the same as in Fig.\,\ref{ug_gz}. 
\label{UB_VK}}
\end{figure}

\begin{figure}
\centering
\resizebox{\hsize}{!}{\includegraphics[clip=true]{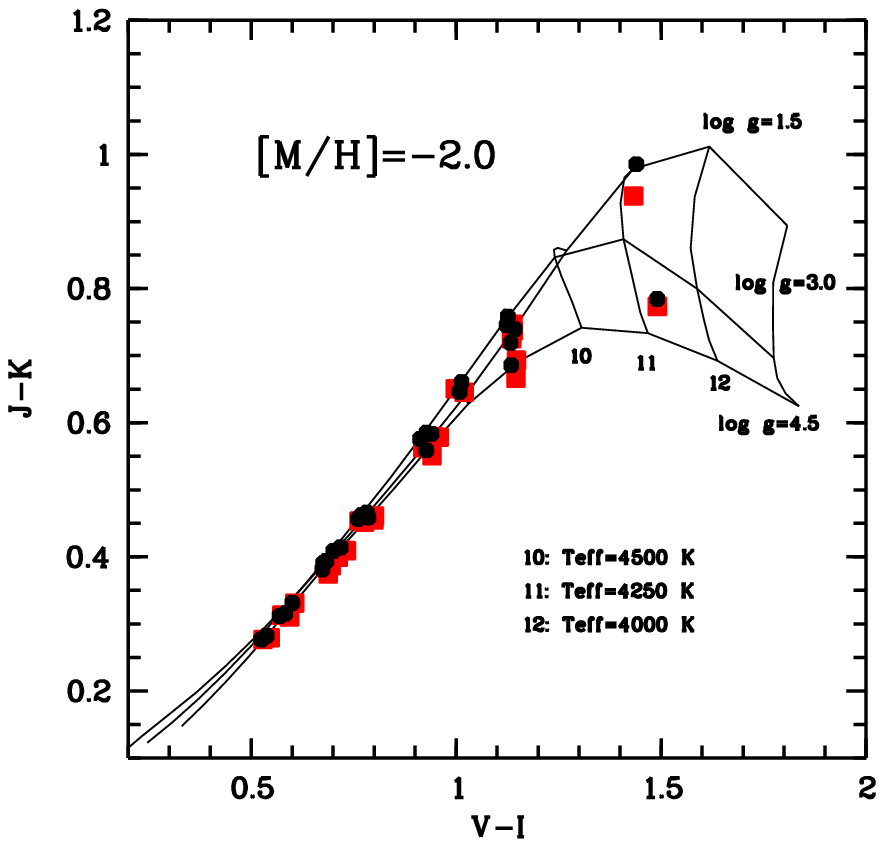}}
\resizebox{\hsize}{!}{\includegraphics[clip=true]{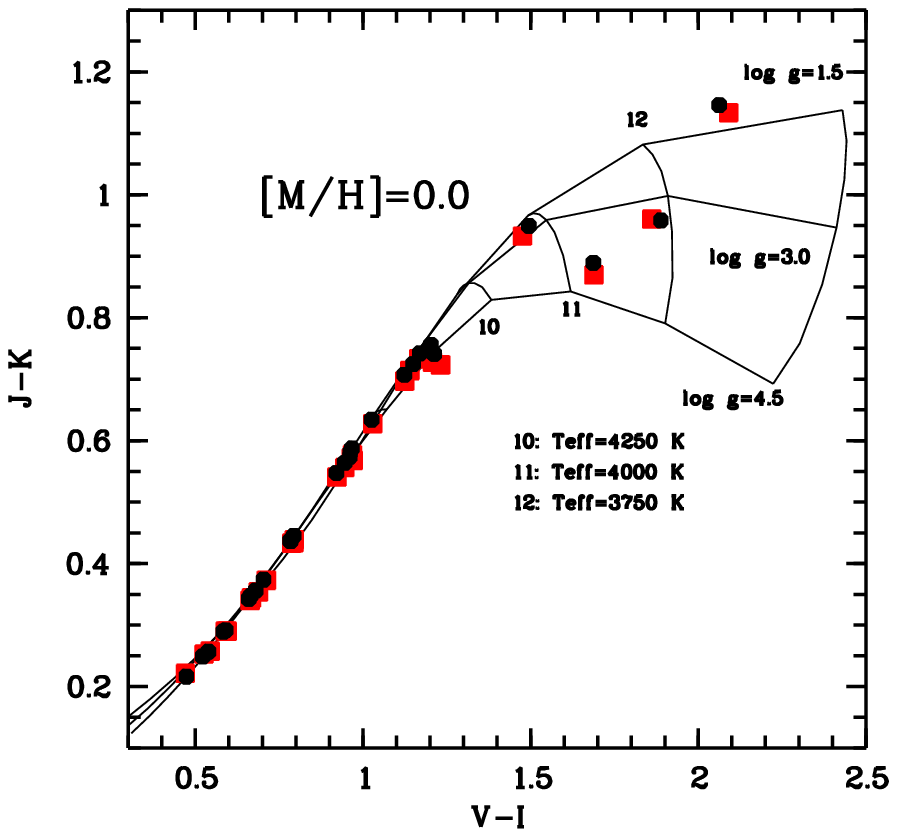}}
\caption{A $J-K$ vs. $V-I$ colour-colour diagram
for solar metallicity (bottom panel) and --2.0 (upper panel). 
The meaning of the lines and symbols is  the same as in Fig.\,\ref{ug_gz}. 
\label{VI_JK}}
\end{figure}

In Fig.\,\ref{UB_VK} we show the $U-B$ versus $V-K$ diagram, which is
morphologically similar to the  $u-g$ versus $g-z$, 
and thus the effect of granulation is similar.
However, the use of the longer baseline colour $V-K$ makes another
effect apparent. At all metallicities, the 3D-corrected colours
are redder for the hotter models and bluer for the cooler
models. This results in a more compressed scale in $V-K$. 
It is also apparent that the difference in inferred temperature
is smaller in this case, less than 100\,K for all gravities and
metallicities. 

In Fig.\,\ref{VI_JK} we compare two colours that are often used
as temperature indicators: $V-I$ and $J-K$.
There is a tight correlation between the two colours, up to about 
4250\,K when the relation splits up and the two colours respond
differently to changes in gravity. The ``3D-corrected'' colours show exactly the
same behaviour, except that the relation is offset by 0.05\,mag.

\begin{figure}
\centering
\resizebox{\hsize}{!}{\includegraphics[clip=true]{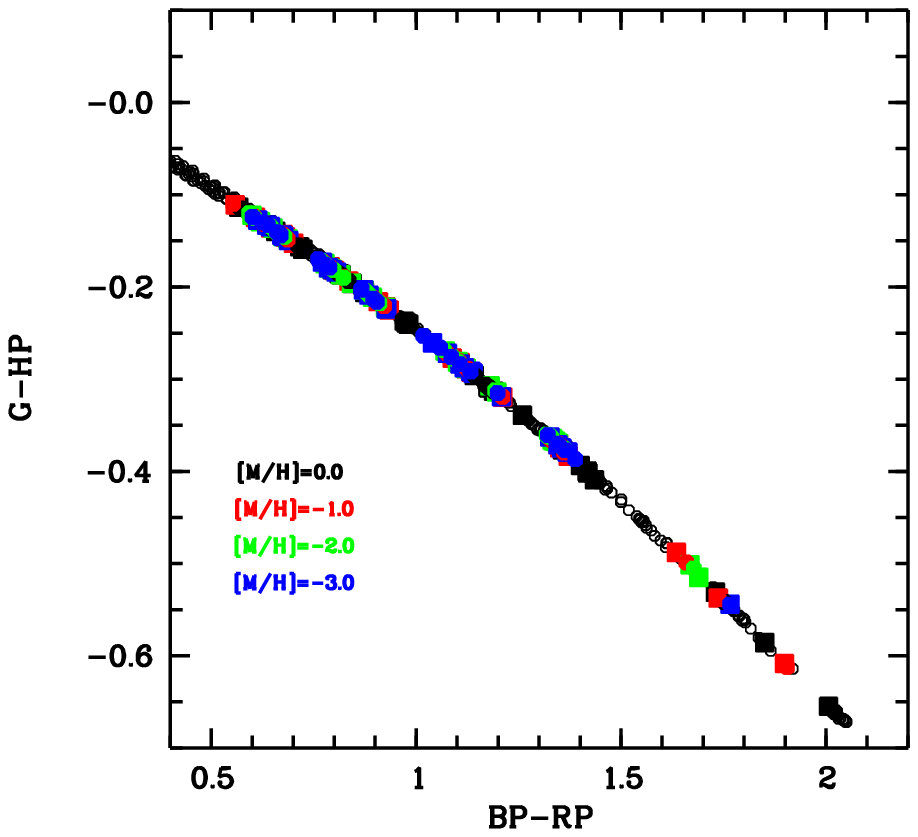}}
\caption{The colour G$-H_p$ as a function of the BP--RP colour.  
Colours computed from models of different metallicity are shown with 
a different colour: black for [M/H]=0.0, red for [M/H]=--1.0,
green for [M/H]=--2.0 and blue for [M/H]=--3.0.
Open circles are the colours computed from ATLAS models, while filled squares
are the corresponding 3D-corrected colours. 
We stress that the Gaia bands are the pre-launch bands.
\label{GHP}}
\end{figure}

\begin{figure}
\centering
\resizebox{\hsize}{!}{\includegraphics[clip=true]{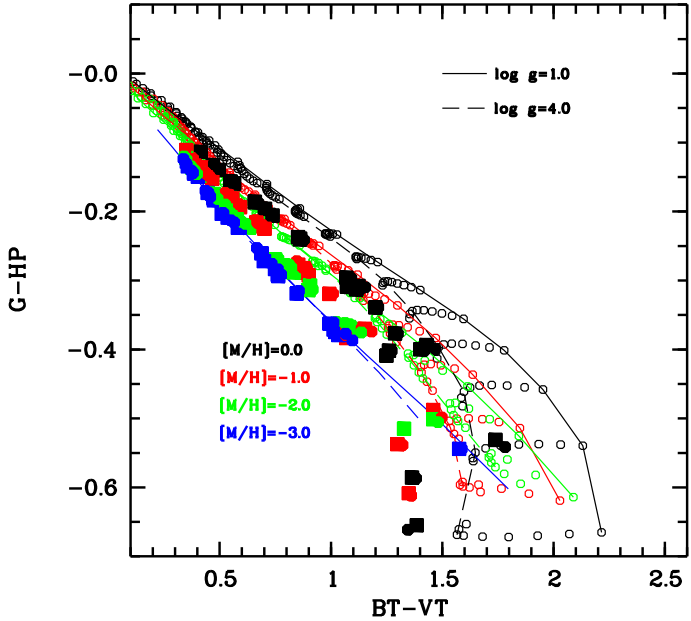}}
\caption{The colour G$-H_p$ as a function of the BT--VT colour. Symbols
are the same as in Fig.\,\ref{GHP}. 
The dashed lines connect the points from the \citet{CK03} grid.
\label{GHPBTVT}}
\end{figure}

\begin{figure}
\centering
\resizebox{\hsize}{!}{\includegraphics[clip=true]{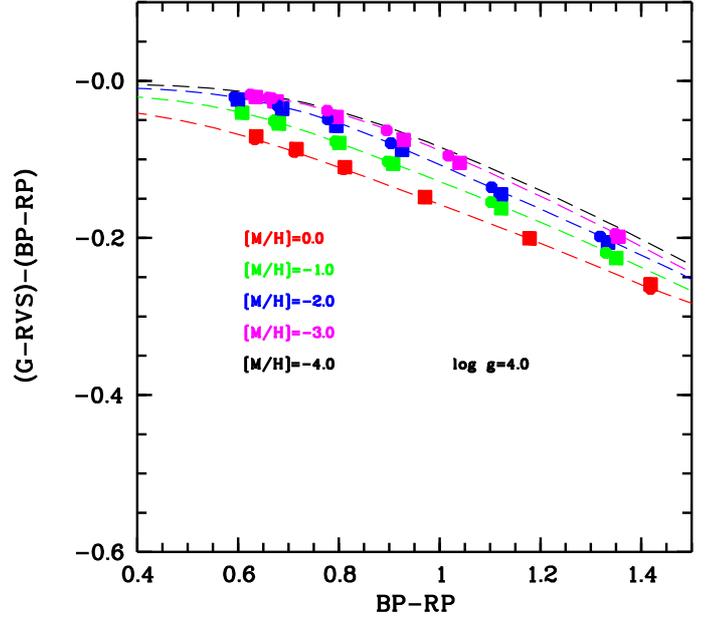}}
\caption{The $CaT$ colour  as a function of the $BP-RP$ colour for models
with log g=4.0. Filled circles are the colours from the ATLAS models and
filled squares are the 3D-corrected colours.  
\label{Ca}}
\end{figure}

Let us turn our attention to the Gaia photometry. We should
warn the reader that the Gaia colours have been computed using the pre-launch
transmission curves. This is useful to gain an understanding of the general
properties of this photometry. Detailed comparison with the observations
will be possible when the post-launch transmission curves become
available, after the Gaia Data Release 2.
We notice that the wide band used for the Gaia photometry, G, is 
sufficiently different from the analogue band of Hipparcos, $H_p$, that the
colour G$-H_p$ is tightly correlated with effective temperature. 
Even more interestingly, it is very tightly correlated with the colour
that can be formed from the Blue Prism (BP) and Red Prism (RP) spectra, BP--RP.
In Fig.\,\ref{GHP} we show how this colour-colour diagram provides a very
tight correlation, essentially putting all the stars of any metallicity
or gravity on the same relation. It is interesting to note that the
3D corrections do not destroy this correlation, since all the points
move along the curve.
This should be a very robust correlation and should prove useful to
determine the effective pass-bands of the Gaia system and reddening and to 
detect binary systems. Any star that is an outlier with respect to such
a correlation should be considered as a potential binary system.
The  G$-H_p$ correlates with any temperature-dependent colour, 
in particular with any colour that measures the slope of
the Paschen continuum. With other colours, however, the relation splits
due to gravity and metallicity dependencies. 
As an example, we show its correlation with the Tycho $B_T - V_T$ colour
in Fig.\,\ref{GHPBTVT}.
In this case the 3D corrections induce  relations that are distinctly different
from what is deduced from the 1D model atmospheres. 
Because of the splitting seen due to 
surface gravity and metallicity, such a colour-colour diagram has a clear
diagnostic potential, however the theoretical colours must be properly
modelled.

Finally let us turn our attention to the potential of the RVS colour as a metallicity diagnostic. 
At the faint end of the sensitivity of the RVS spectrograph \citep{RVS},
in the interval G=15 to G=17, it is likely that the spectra will
be too noisy to allow a metallicity determination. Yet, since the spectra
will be flux-calibrated one can derive an RVS magnitude, that is, a narrow-band filter centered on the IR \ion{Ca}{ii} triplet. In ground-based
observations, such a filter is not very useful, since the signal of
the \ion{Ca}{ii} lines is severely diluted by the signal coming from the
atmospheric OH emission lines. Data taken from space, however, like Gaia, do
not suffer from such a limitation.
In Fig.\,\ref{Ca} we show the behaviour of the $CaT$ colour
as a function of the $BP-RP$ colour. 
We show several metallicities for log g=4.0 and it is  clear that even
a precision of 0.02 mag in this colour should allow an estimate
of the metallicity with a precision of 0.5\, dex, down to a metallicity of --3.0, 
at least. The colour is not strongly sensitive to gravity. However
Gaia will also provide the parallaxes, thus the surface gravity will be
known. It is interesting to see that the 3D corrections move the
points along the curve defined by the 1D models, so that the
metallicity sensitivity of the colour remains the same.
The metallicity information from the RVS magnitude will be
complementary to that coming from the full spectral energy distribution,
derived from the prism spectra, for stars for which both will be available.

\section{Conclusions}

We have used the CIFIST grid of \cobold\
models to investigate the effects of granulation
on fluxes and colours of stars of spectral type F, G, and K.
Our investigation is exploratory, since we realise that treating
scattering as true absorption, as done in the current
version of the CIFIST models, leads to incorrect results in the
computed emerging fluxes, especially in the blue and UV regions.  
The influence of this assumption can be quantified in terms of
colours and is of the order of 0.02 magnitudes for colours
involving blue or UV bands. 
While the next generation of the CIFIST grid is being computed
using the approximate treatment of scattering suggested
by \citet{Hayek}, we believe it is still interesting to 
publish the results obtained from the current version
of the CIFIST grid.

The effects of granulation on colours are generally small, 
of a few hundredths of a magnitude; they cannot, however be
considered negligible. On some colours, such as $g-z,$ the 
effect can translate to a temperature estimate that 
is hotter by 200\,K. 

We publish tables with 3D corrections that can be applied to 
colours computed from any 1D model atmosphere. For  \teff $\ge 5000$K,
the corrections are smooth enough, as a function of atmospheric parameters,
that it is possible to interpolate the corrections between
grid points; thus the coarseness of the CIFIST grid should
not be a major limitation. However at the cool end there are still
far too few models to allow a reliable interpolation.

We have investigated the effects on the Gaia photometric system, 
although only the pre-launch transmission curves are available. 
We confirm that a tight correlation between the G$-H_p$ colour
and the BP--RP colour should exist, and that this correlation
should be the same also after granulation effects are taken
into account.
We have also investigated the potential of the
(G--RVS)--(BP--RP) colour as a metallicity diagnostic and we confirm
that it is indeed a good indicator, even after 3D corrections have been
taken into account, provided this colour can be measured with 
a precision of 0.02\,mag or better.  If one is only interested
in selecting stars that are more metal-poor than --3.0, the colour
is still very powerful even if the precision is of the order
of 0.05\, mag.

\begin{acknowledgements}
We are grateful to the referee, Santi Cassisi, whose
report helped us to improve the paper.
This project has been supported by fondation MERAC and
Observatoire de Paris. We acknowledge financial support from 
CNRS Institut National de Sciences de l'Univers Programme
National de Cosmologie et Galaxies and Programme National de Physique 
Stellaire.  This work was granted access  to  the  
HPC  resources  of  MesoPSL  financed
by   the   R\'egion   \^{I}le   de   France   and   
the   project Equip@Meso (reference ANR-10-EQPX-29-01) of
the programme Investissements d'Avenir supervised
by  the  Agence  Nationale  pour  la  Recherche. 
HGL and DH acknowledge financial support by the Sonderforschungsbereich SFB\,881
``The Milky Way System'' (subproject A4) of the German Research Foundation
(DFG). This research was supported by a grant (MIP-089/2015) from the Research Council of Lithuania.
\end{acknowledgements}

\bibliographystyle{aa}

\begin{appendix}

\section{Tables of colours and colour corrections.}

In this Appendix we present the 3D-1D colour corrections for colours
in the Johnson-Cousins, 2MASS, Hipparcos, Gaia and SDSS systems.
Bolometric corrections are also provided for the Gaia $G$ magnitude.
For each colour, we provide the colour from the ATLAS model, and the ``3D-corrected'' colour, that is, the ATLAS colour to which we added
the corresponding 3D-1D corrections. 
The 3D-corrected colours can be used directly to interpret observed colours. 
The 3D corrections can also be used to correct colours computed from any
1D model atmosphere.  

\begin{table*}
\caption{\label{ubvrim00}
Colours and corrections for the Johnson-Cousins system for metallicity [M/H]=0.0. In columns labelled  `ATLAS$_c$' the 3D correction
has been added to the ATLAS colour. }
\renewcommand{\tabcolsep}{3pt}
\tabskip=0pt
\begin{center}
\begin{tabular}{llllllllllllll}
\hline\noalign{\smallskip}
\multicolumn{1}{c}{\teff} & 
\multicolumn{1}{c}{log g} & 
\multicolumn{1}{c}{$U-B$} &
\multicolumn{1}{c}{$U-B$} &
\multicolumn{1}{c}{$\Delta_{U-B}$} &
\multicolumn{1}{c}{$B-V$} &
\multicolumn{1}{c}{$B-V$} &
\multicolumn{1}{c}{$\Delta_{B-V}$} &
\multicolumn{1}{c}{$V-R$} &
\multicolumn{1}{c}{$V-R$} &
\multicolumn{1}{c}{$\Delta_{V-R}$} &
\multicolumn{1}{c}{$V-I$} &
\multicolumn{1}{c}{$V-I$} &
\multicolumn{1}{c}{$\Delta_{V-I}$} \\
\multicolumn{2}{c}{ } &  
\multicolumn{1}{c}{ATLAS} &
\multicolumn{1}{c}{ATLAS$_c$}&
\multicolumn{1}{c}{3D-1D}&
\multicolumn{1}{c}{ATLAS} &
\multicolumn{1}{c}{ATLAS$_c$}&
\multicolumn{1}{c}{3D-1D}&
\multicolumn{1}{c}{ATLAS} &
\multicolumn{1}{c}{ATLAS$_c$}&
\multicolumn{1}{c}{3D-1D}&
\multicolumn{1}{c}{ATLAS} &
\multicolumn{1}{c}{ATLAS$_c$}&
\multicolumn{1}{c}{3D-1D} \\
\hline\noalign{\smallskip}
\hline\noalign{\smallskip}
3635  &1.00 & $+1.880$ & $+1.792$ &$ -0.089$ &$ +1.671$ &$ +1.583$ &$ -0.014$ &$ +0.944$ &$ +0.956$ &$ +0.012$ &$ +2.062$&$ +2.090$&$  +0.028$ \\
3813  &4.00 & $+1.123$ & $+1.073$ &$ -0.050$ &$ +1.306$ &$ +1.256$ &$ +0.032$ &$ +0.878$ &$ +0.872$ &$ -0.006$ &$ +1.888$&$ +1.861$&$  -0.027$ \\
4018  &1.50 & $+1.688$ & $+1.580$ &$ -0.108$ &$ +1.541$ &$ +1.433$ &$ -0.034$ &$ +0.783$ &$ +0.764$ &$ -0.019$ &$ +1.495$&$ +1.476$&$  -0.019$ \\
3964  &4.50 & $+1.095$ & $+1.035$ &$ -0.060$ &$ +1.314$ &$ +1.255$ &$ -0.015$ &$ +0.844$ &$ +0.832$ &$ -0.012$ &$ +1.687$&$ +1.689$&$  +0.002$ \\
4501  &2.00 & $+1.069$ & $+1.011$ &$ -0.058$ &$ +1.268$ &$ +1.210$ &$ -0.027$ &$ +0.608$ &$ +0.601$ &$ -0.007$ &$ +1.149$&$ +1.140$&$  -0.009$ \\
4477  &2.50 & $+1.004$ & $+0.979$ &$ -0.025$ &$ +1.235$ &$ +1.210$ &$ -0.012$ &$ +0.618$ &$ +0.615$ &$ -0.003$ &$ +1.169$&$ +1.167$&$  -0.003$ \\
\end{tabular}
\end{center}
\end{table*}

\begin{table*}
\caption{\label{ubvrim10}
Colours and corrections for the Johnson-Cousins system for metallicity [M/H]=--1.0. In columns labelled `ATLAS$_c$' the 3D correction
has been added to the ATLAS colour. }
\renewcommand{\tabcolsep}{3pt}
\tabskip=0pt
\begin{center}
\begin{tabular}{llllllllllllll}
\hline\noalign{\smallskip}
\multicolumn{1}{c}{\teff} & 
\multicolumn{1}{c}{log g} & 
\multicolumn{1}{c}{$U-B$} &
\multicolumn{1}{c}{$U-B$} &
\multicolumn{1}{c}{$\Delta_{U-B}$} &
\multicolumn{1}{c}{$B-V$} &
\multicolumn{1}{c}{$B-V$} &
\multicolumn{1}{c}{$\Delta_{B-V}$} &
\multicolumn{1}{c}{$V-R$} &
\multicolumn{1}{c}{$V-R$} &
\multicolumn{1}{c}{$\Delta_{V-R}$} &
\multicolumn{1}{c}{$V-I$} &
\multicolumn{1}{c}{$V-I$} &
\multicolumn{1}{c}{$\Delta_{V-I}$} \\
\multicolumn{2}{c}{ } &  
\multicolumn{1}{c}{ATLAS} &
\multicolumn{1}{c}{ATLAS$_c$}&
\multicolumn{1}{c}{3D-1D}&
\multicolumn{1}{c}{ATLAS} &
\multicolumn{1}{c}{ATLAS$_c$}&
\multicolumn{1}{c}{3D-1D}&
\multicolumn{1}{c}{ATLAS} &
\multicolumn{1}{c}{ATLAS$_c$}&
\multicolumn{1}{c}{3D-1D}&
\multicolumn{1}{c}{ATLAS} &
\multicolumn{1}{c}{ATLAS$_c$}&
\multicolumn{1}{c}{3D-1D} \\
\hline\noalign{\smallskip}
\hline\noalign{\smallskip}
3790  &4.00 &$ +0.882$ &$ +0.843$ &$ -0.038$ &$ +1.326$ &$ +1.288$ &$ -0.012$ &$ +0.873$ &$ +0.859$ &$ -0.013$ &$ +1.744$ &$ +1.739$ &$ -0.005$\\
4040  &1.50 &$ +1.068$ &$ +0.996$ &$ -0.073$ &$ +1.339$ &$ +1.266$ &$ -0.031$ &$ +0.736$ &$ +0.718$ &$ -0.018$ &$ +1.424$ &$ +1.403$ &$ -0.021$\\
4001  &4.50 &$ +0.800$ &$ +0.768$ &$ -0.032$ &$ +1.268$ &$ +1.236$ &$ -0.020$ &$ +0.818$ &$ +0.806$ &$ -0.012$ &$ +1.555$ &$ +1.560$ &$ +0.006$\\
4485  &2.00 &$ +0.553$ &$ +0.526$ &$ -0.027$ &$ +1.069$ &$ +1.042$ &$ -0.021$ &$ +0.575$ &$ +0.571$ &$ -0.004$ &$ +1.137$ &$ +1.132$ &$ -0.005$\\
4490  &2.50 &$ +0.484$ &$ +0.479$ &$ -0.004$ &$ +1.036$ &$ +1.031$ &$ -0.005$ &$ +0.576$ &$ +0.577$ &$ +0.001$ &$ +1.136$ &$ +1.139$ &$ +0.003$\\
4525  &4.00 &$ +0.417$ &$ +0.440$ &$ +0.023$ &$ +0.978$ &$ +1.000$ &$ +0.014$ &$ +0.595$ &$ +0.609$ &$ +0.013$ &$ +1.145$ &$ +1.167$ &$ +0.022$\\
\end{tabular}
\end{center}
\end{table*}

\begin{table*}
\caption{\label{ubvrim20}
Colours and corrections for the Johnson-Cousins system for metallicity [M/H]=--2.0. In columns labelled `ATLAS$_c$' the 3D correction
has been added to the ATLAS colour. }
\renewcommand{\tabcolsep}{3pt}
\tabskip=0pt
\begin{center}
\begin{tabular}{llllllllllllll}
\hline\noalign{\smallskip}
\multicolumn{1}{c}{\teff} & 
\multicolumn{1}{c}{log g} & 
\multicolumn{1}{c}{$U-B$} &
\multicolumn{1}{c}{$U-B$} &
\multicolumn{1}{c}{$\Delta_{U-B}$} &
\multicolumn{1}{c}{$B-V$} &
\multicolumn{1}{c}{$B-V$} &
\multicolumn{1}{c}{$\Delta_{B-V}$} &
\multicolumn{1}{c}{$V-R$} &
\multicolumn{1}{c}{$V-R$} &
\multicolumn{1}{c}{$\Delta_{V-R}$} &
\multicolumn{1}{c}{$V-I$} &
\multicolumn{1}{c}{$V-I$} &
\multicolumn{1}{c}{$\Delta_{V-I}$} \\
\multicolumn{2}{c}{ } &  
\multicolumn{1}{c}{ATLAS} &
\multicolumn{1}{c}{ATLAS$_c$}&
\multicolumn{1}{c}{3D-1D}&
\multicolumn{1}{c}{ATLAS} &
\multicolumn{1}{c}{ATLAS$_c$}&
\multicolumn{1}{c}{3D-1D}&
\multicolumn{1}{c}{ATLAS} &
\multicolumn{1}{c}{ATLAS$_c$}&
\multicolumn{1}{c}{3D-1D}&
\multicolumn{1}{c}{ATLAS} &
\multicolumn{1}{c}{ATLAS$_c$}&
\multicolumn{1}{c}{3D-1D} \\
\hline\noalign{\smallskip}
\hline\noalign{\smallskip}
4001  &1.50& $ +0.895$&$  +0.857$&$  -0.038$&$  +1.336$&$  +1.298$&$  -0.015$&$  +0.733$&$  +0.729$&$  -0.004$&$  +1.440$&$  +1.433$&$  -0.007$\\
4000  &4.50& $ +0.657$&$  +0.654$&$  -0.003$&$  +1.255$&$  +1.252$&$  -0.001$&$  +0.785$&$  +0.784$&$  -0.001$&$  +1.490$&$  +1.491$&$  +0.000$\\
4506  &1.50& $ +0.440$&$  +0.417$&$  -0.023$&$  +1.050$&$  +1.027$&$  -0.039$&$  +0.568$&$  +0.563$&$  -0.004$&$  +1.143$&$  +1.139$&$  -0.004$\\
4506  &2.00& $ +0.301$&$  +0.300$&$  -0.000$&$  +0.989$&$  +0.988$&$  -0.011$&$  +0.557$&$  +0.564$&$  +0.007$&$  +1.126$&$  +1.138$&$  +0.013$\\
4478  &2.50& $ +0.344$&$  +0.350$&$  +0.006$&$  +1.006$&$  +1.012$&$  +0.001$&$  +0.555$&$  +0.565$&$  +0.010$&$  +1.123$&$  +1.136$&$  +0.013$\\
4504  &4.00& $ +0.272$&$  +0.283$&$  +0.011$&$  +0.960$&$  +0.971$&$  +0.015$&$  +0.571$&$  +0.582$&$  +0.011$&$  +1.133$&$  +1.147$&$  +0.015$\\
\end{tabular}
\end{center}
\end{table*}

\begin{table*}
\caption{\label{ubvrim30}
Colours and corrections for the Johnson-Cousins system for metallicity [M/H]=--3.0. In columns labelled `ATLAS$_c$' the 3D correction
has been added to the ATLAS colour. }
\renewcommand{\tabcolsep}{3pt}
\tabskip=0pt
\begin{center}
\begin{tabular}{llllllllllllll}
\hline\noalign{\smallskip}
\multicolumn{1}{c}{\teff} & 
\multicolumn{1}{c}{log g} & 
\multicolumn{1}{c}{$U-B$} &
\multicolumn{1}{c}{$U-B$} &
\multicolumn{1}{c}{$\Delta_{U-B}$} &
\multicolumn{1}{c}{$B-V$} &
\multicolumn{1}{c}{$B-V$} &
\multicolumn{1}{c}{$\Delta_{B-V}$} &
\multicolumn{1}{c}{$V-R$} &
\multicolumn{1}{c}{$V-R$} &
\multicolumn{1}{c}{$\Delta_{V-R}$} &
\multicolumn{1}{c}{$V-I$} &
\multicolumn{1}{c}{$V-I$} &
\multicolumn{1}{c}{$\Delta_{V-I}$} \\
\multicolumn{2}{c}{ } &  
\multicolumn{1}{c}{ATLAS} &
\multicolumn{1}{c}{ATLAS$_c$}&
\multicolumn{1}{c}{3D-1D}&
\multicolumn{1}{c}{ATLAS} &
\multicolumn{1}{c}{ATLAS$_c$}&
\multicolumn{1}{c}{3D-1D}&
\multicolumn{1}{c}{ATLAS} &
\multicolumn{1}{c}{ATLAS$_c$}&
\multicolumn{1}{c}{3D-1D}&
\multicolumn{1}{c}{ATLAS} &
\multicolumn{1}{c}{ATLAS$_c$}&
\multicolumn{1}{c}{3D-1D} \\
\hline\noalign{\smallskip}
\hline\noalign{\smallskip}
3990  &1.50& $ +0.962$&$  +0.980$&$  +0.018$&$  +1.437$&$  +1.456$&$  -0.010$&$  +0.769$&$  +0.771$&$  +0.002$&$  +1.515$&$  +1.518$&$  +0.003$\\
4503  &1.50& $ +0.373$&$  +0.338$&$  -0.035$&$  +1.039$&$  +1.004$&$  -0.050$&$  +0.583$&$  +0.576$&$  -0.006$&$  +1.177$&$  +1.171$&$  -0.007$\\
4488  &2.00& $ +0.302$&$  +0.298$&$  -0.005$&$  +1.011$&$  +1.006$&$  -0.016$&$  +0.571$&$  +0.577$&$  +0.006$&$  +1.159$&$  +1.170$&$  +0.011$\\
4523  &2.50& $ +0.214$&$  +0.207$&$  -0.006$&$  +0.965$&$  +0.958$&$  -0.012$&$  +0.553$&$  +0.560$&$  +0.007$&$  +1.126$&$  +1.136$&$  +0.011$\\
4494  &4.00& $ +0.193$&$  +0.201$&$  +0.008$&$  +0.973$&$  +0.981$&$  +0.005$&$  +0.572$&$  +0.578$&$  +0.005$&$  +1.157$&$  +1.164$&$  +0.007$\\
4522  &4.50& $ +0.186$&$  +0.193$&$  +0.007$&$  +0.967$&$  +0.974$&$  +0.003$&$  +0.571$&$  +0.575$&$  +0.004$&$  +1.152$&$  +1.156$&$  +0.004$\\
\end{tabular}
\end{center}
\end{table*}

\begin{table*}
\caption{\label{2masshipm00}
Colours and corrections for the 2MASS and Hipparcos-Tycho systems for metallicity [M/H]=0.0. In columns labelled `ATLAS$_c$' the 3D correction
has been added to the ATLAS colour. }
\renewcommand{\tabcolsep}{3pt}
\tabskip=0pt
\begin{center}
\begin{tabular}{llllllllllllll}
\hline\noalign{\smallskip}
\multicolumn{1}{c}{\teff} & 
\multicolumn{1}{c}{log g} & 
\multicolumn{1}{c}{$J-K$} &
\multicolumn{1}{c}{$J-K$} &
\multicolumn{1}{c}{$\Delta_{J-K}$} &
\multicolumn{1}{c}{$V-K$} &
\multicolumn{1}{c}{$V-K$} &
\multicolumn{1}{c}{$\Delta_{V-K}$} &
\multicolumn{1}{c}{$V-H_p$} &
\multicolumn{1}{c}{$V-H_p$} &
\multicolumn{1}{c}{$\Delta_{V-H_p}$} &
\multicolumn{1}{c}{$B_T-V_T$} &
\multicolumn{1}{c}{$B_T-V_T$} &
\multicolumn{1}{c}{$\Delta_{B_T-V_T}$} \\
\multicolumn{2}{c}{ } &  
\multicolumn{1}{c}{ATLAS} &
\multicolumn{1}{c}{ATLAS$_c$}&
\multicolumn{1}{c}{3D-1D}&
\multicolumn{1}{c}{ATLAS} &
\multicolumn{1}{c}{ATLAS$_c$}&
\multicolumn{1}{c}{3D-1D}&
\multicolumn{1}{c}{ATLAS} &
\multicolumn{1}{c}{ATLAS$_c$}&
\multicolumn{1}{c}{3D-1D}&
\multicolumn{1}{c}{ATLAS} &
\multicolumn{1}{c}{ATLAS$_c$}&
\multicolumn{1}{c}{3D-1D} \\
\hline\noalign{\smallskip}
\hline\noalign{\smallskip}
3635  &1.00 &$ +1.146$ &$ +1.134$ &$ -0.012$ &$ +4.567$ &$ +4.637$ &$ +0.070$ &$ -0.179$ &$ -0.172$ &$ +0.007$ &$ +1.857$ &$ +1.840$ &$ -0.017$\\
3813  &4.00 &$ +0.958$ &$ +0.960$ &$ +0.002$ &$ +4.026$ &$ +3.964$ &$ -0.061$ &$ -0.150$ &$ -0.159$ &$ -0.009$ &$ +1.346$ &$ +1.385$ &$ +0.039$\\
4018  &1.50 &$ +0.949$ &$ +0.933$ &$ -0.017$ &$ +3.513$ &$ +3.463$ &$ -0.050$ &$ -0.217$ &$ -0.218$ &$ -0.001$ &$ +1.782$ &$ +1.739$ &$ -0.043$\\
3964  &4.50 &$ +0.889$ &$ +0.870$ &$ -0.020$ &$ +3.627$ &$ +3.594$ &$ -0.033$ &$ -0.162$ &$ -0.164$ &$ -0.002$ &$ +1.382$ &$ +1.364$ &$ -0.018$\\
4501  &2.00 &$ +0.725$ &$ +0.714$ &$ -0.011$ &$ +2.717$ &$ +2.681$ &$ -0.036$ &$ -0.225$ &$ -0.222$ &$ +0.003$ &$ +1.469$ &$ +1.430$ &$ -0.039$\\
4477  &2.50 &$ +0.742$ &$ +0.733$ &$ -0.009$ &$ +2.762$ &$ +2.741$ &$ -0.021$ &$ -0.215$ &$ -0.214$ &$ +0.001$ &$ +1.419$ &$ +1.402$ &$ -0.017$\\
\end{tabular}
\end{center}
\end{table*}

\begin{table*}
\caption{\label{2masshipm10}
Colours and corrections for the 2MASS and Hipparcos-Tycho systems for metallicity [M/H]=--1.0. In columns labelled `ATLAS$_c$' the 3D correction
has been added to the ATLAS colour. }
\renewcommand{\tabcolsep}{3pt}
\tabskip=0pt
\begin{center}
\begin{tabular}{llllllllllllll}
\hline\noalign{\smallskip}
\multicolumn{1}{c}{\teff} & 
\multicolumn{1}{c}{log g} & 
\multicolumn{1}{c}{$J-K$} &
\multicolumn{1}{c}{$J-K$} &
\multicolumn{1}{c}{$\Delta_{J-K}$} &
\multicolumn{1}{c}{$V-K$} &
\multicolumn{1}{c}{$V-K$} &
\multicolumn{1}{c}{$\Delta_{V-K}$} &
\multicolumn{1}{c}{$V-H_p$} &
\multicolumn{1}{c}{$V-H_p$} &
\multicolumn{1}{c}{$\Delta_{V-H_p}$} &
\multicolumn{1}{c}{$B_T-V_T$} &
\multicolumn{1}{c}{$B_T-V_T$} &
\multicolumn{1}{c}{$\Delta_{B_T-V_T}$} \\
\multicolumn{2}{c}{ } &  
\multicolumn{1}{c}{ATLAS} &
\multicolumn{1}{c}{ATLAS$_c$}&
\multicolumn{1}{c}{3D-1D}&
\multicolumn{1}{c}{ATLAS} &
\multicolumn{1}{c}{ATLAS$_c$}&
\multicolumn{1}{c}{3D-1D}&
\multicolumn{1}{c}{ATLAS} &
\multicolumn{1}{c}{ATLAS$_c$}&
\multicolumn{1}{c}{3D-1D}&
\multicolumn{1}{c}{ATLAS} &
\multicolumn{1}{c}{ATLAS$_c$}&
\multicolumn{1}{c}{3D-1D} \\
\hline\noalign{\smallskip}
\hline\noalign{\smallskip}
3790  &4.00&  $+0.859$ &$ +0.842$ &$ -0.017$ &$ +3.683$ &$ +3.649$ &$ -0.034$ &$ -0.157$ &$ -0.160$ &$ -0.003$ &$ +1.360$ &$ +1.350$  &$-0.011$\\
4040  &1.50&  $+0.968$ &$ +0.940$ &$ -0.028$ &$ +3.430$ &$ +3.358$ &$ -0.072$ &$ -0.199$ &$ -0.200$ &$ -0.001$ &$ +1.498$ &$ +1.460$  &$-0.039$\\
4001  &4.50&  $+0.820$ &$ +0.802$ &$ -0.018$ &$ +3.345$ &$ +3.329$ &$ -0.017$ &$ -0.163$ &$ -0.163$ &$ -0.001$ &$ +1.322$ &$ +1.298$  &$-0.023$\\
4485  &2.00&  $+0.756$ &$ +0.742$ &$ -0.014$ &$ +2.734$ &$ +2.700$ &$ -0.034$ &$ -0.195$ &$ -0.192$ &$ +0.003$ &$ +1.182$ &$ +1.153$  &$-0.029$\\
4490  &2.50&  $+0.758$ &$ +0.741$ &$ -0.017$ &$ +2.733$ &$ +2.707$ &$ -0.026$ &$ -0.186$ &$ -0.185$ &$ +0.001$ &$ +1.138$ &$ +1.128$  &$-0.010$\\
4525  &4.00&  $+0.743$ &$ +0.728$ &$ -0.015$ &$ +2.710$ &$ +2.712$ &$ +0.002$ &$ -0.165$ &$ -0.165$ &$ +0.000$ &$ +1.057$ &$ +1.068$  &$+0.011$\\
\end{tabular}
\end{center}
\end{table*}

\begin{table*}
\caption{\label{2masshipm20}
Colours and corrections for the 2MASS and Hipparcos-Tycho systems for metallicity [M/H]=--2.0. In columns labelled `ATLAS$_c$' the 3D correction
has been added to the ATLAS colour. }
\renewcommand{\tabcolsep}{3pt}
\tabskip=0pt
\begin{center}
\begin{tabular}{llllllllllllll}
\hline\noalign{\smallskip}
\multicolumn{1}{c}{\teff} & 
\multicolumn{1}{c}{log g} & 
\multicolumn{1}{c}{$J-K$} &
\multicolumn{1}{c}{$J-K$} &
\multicolumn{1}{c}{$\Delta_{J-K}$} &
\multicolumn{1}{c}{$V-K$} &
\multicolumn{1}{c}{$V-K$} &
\multicolumn{1}{c}{$\Delta_{V-K}$} &
\multicolumn{1}{c}{$V-H_p$} &
\multicolumn{1}{c}{$V-H_p$} &
\multicolumn{1}{c}{$\Delta_{V-H_p}$} &
\multicolumn{1}{c}{$B_T-V_T$} &
\multicolumn{1}{c}{$B_T-V_T$} &
\multicolumn{1}{c}{$\Delta_{B_T-V_T}$} \\
\multicolumn{2}{c}{ } &  
\multicolumn{1}{c}{ATLAS} &
\multicolumn{1}{c}{ATLAS$_c$}&
\multicolumn{1}{c}{3D-1D}&
\multicolumn{1}{c}{ATLAS} &
\multicolumn{1}{c}{ATLAS$_c$}&
\multicolumn{1}{c}{3D-1D}&
\multicolumn{1}{c}{ATLAS} &
\multicolumn{1}{c}{ATLAS$_c$}&
\multicolumn{1}{c}{3D-1D}&
\multicolumn{1}{c}{ATLAS} &
\multicolumn{1}{c}{ATLAS$_c$}&
\multicolumn{1}{c}{3D-1D} \\
\hline\noalign{\smallskip}
\hline\noalign{\smallskip}
4001  &1.50 & $+0.985$ &$ +0.938$ &$ -0.047$ &$ +3.479$ &$ +3.395$ &$ -0.084$ &$ -0.202$ &$ -0.201$ &$ +0.001$ &$ +1.481$ &$ +1.460$ &$ -0.022$\\
4000  &4.50 & $+0.785$ &$ +0.773$ &$ -0.011$ &$ +3.216$ &$ +3.201$ &$ -0.015$ &$ -0.170$ &$ -0.170$ &$ -0.000$ &$ +1.330$ &$ +1.329$ &$ -0.001$\\
4506  &1.50 & $+0.740$ &$ +0.737$ &$ -0.003$ &$ +2.711$ &$ +2.690$ &$ -0.021$ &$ -0.197$ &$ -0.190$ &$ +0.007$ &$ +1.136$ &$ +1.086$ &$ -0.050$\\
4506  &2.00 & $+0.758$ &$ +0.747$ &$ -0.012$ &$ +2.723$ &$ +2.711$ &$ -0.012$ &$ -0.184$ &$ -0.180$ &$ +0.004$ &$ +1.063$ &$ +1.045$ &$ -0.017$\\
4478  &2.50 & $+0.745$ &$ +0.725$ &$ -0.020$ &$ +2.697$ &$ +2.676$ &$ -0.021$ &$ -0.190$ &$ -0.187$ &$ +0.003$ &$ +1.083$ &$ +1.080$ &$ -0.003$\\
4504  &4.00 & $+0.719$ &$ +0.693$ &$ -0.026$ &$ +2.661$ &$ +2.635$ &$ -0.025$ &$ -0.171$ &$ -0.172$ &$ -0.001$ &$ +1.025$ &$ +1.040$ &$ +0.015$\\
\end{tabular}
\end{center}
\end{table*}

\begin{table*}
\caption{\label{2masshipm30}
Colours and corrections for the 2MASS and Hipparcos-Tycho systems for metallicity [M/H]=--3.0. In columns labelled `ATLAS$_c$' the 3D correction
has been added to the ATLAS colour. }
\renewcommand{\tabcolsep}{3pt}
\tabskip=0pt
\begin{center}
\begin{tabular}{llllllllllllll}
\hline\noalign{\smallskip}
\multicolumn{1}{c}{\teff} & 
\multicolumn{1}{c}{log g} & 
\multicolumn{1}{c}{$J-K$} &
\multicolumn{1}{c}{$J-K$} &
\multicolumn{1}{c}{$\Delta_{J-K}$} &
\multicolumn{1}{c}{$V-K$} &
\multicolumn{1}{c}{$V-K$} &
\multicolumn{1}{c}{$\Delta_{V-K}$} &
\multicolumn{1}{c}{$V-H_p$} &
\multicolumn{1}{c}{$V-H_p$} &
\multicolumn{1}{c}{$\Delta_{V-H_p}$} &
\multicolumn{1}{c}{$B_T-V_T$} &
\multicolumn{1}{c}{$B_T-V_T$} &
\multicolumn{1}{c}{$\Delta_{B_T-V_T}$} \\
\multicolumn{2}{c}{ } &  
\multicolumn{1}{c}{ATLAS} &
\multicolumn{1}{c}{ATLAS$_c$}&
\multicolumn{1}{c}{3D-1D}&
\multicolumn{1}{c}{ATLAS} &
\multicolumn{1}{c}{ATLAS$_c$}&
\multicolumn{1}{c}{3D-1D}&
\multicolumn{1}{c}{ATLAS} &
\multicolumn{1}{c}{ATLAS$_c$}&
\multicolumn{1}{c}{3D-1D}&
\multicolumn{1}{c}{ATLAS} &
\multicolumn{1}{c}{ATLAS$_c$}&
\multicolumn{1}{c}{3D-1D} \\
\hline\noalign{\smallskip}
\hline\noalign{\smallskip}
3990  &1.50 &$ +0.940$ &$ +0.922$ &$ -0.018$ &$ +3.506$ &$ +3.469$ &$ -0.038$ &$ -0.212$ &$ -0.210$ &$ +0.002$ &$ +1.588$ &$ +1.576$ &$ -0.012$\\
4503  &1.50 &$ +0.725$ &$ +0.727$ &$ +0.002$ &$ +2.728$ &$ +2.709$ &$ -0.019$ &$ -0.195$ &$ -0.186$ &$ +0.008$ &$ +1.096$ &$ +1.033$ &$ -0.063$\\
4488  &2.00 &$ +0.729$ &$ +0.723$ &$ -0.006$ &$ +2.714$ &$ +2.708$ &$ -0.006$ &$ -0.190$ &$ -0.185$ &$ +0.005$ &$ +1.066$ &$ +1.043$ &$ -0.022$\\
4523  &2.50 &$ +0.711$ &$ +0.695$ &$ -0.015$ &$ +2.644$ &$ +2.626$ &$ -0.019$ &$ -0.184$ &$ -0.179$ &$ +0.005$ &$ +1.012$ &$ +0.994$ &$ -0.018$\\
4494  &4.00 &$ +0.660$ &$ +0.654$ &$ -0.006$ &$ +2.589$ &$ +2.585$ &$ -0.004$ &$ -0.180$ &$ -0.180$ &$ +0.001$ &$ +1.015$ &$ +1.020$ &$ +0.005$\\
4522  &4.50 &$ +0.634$ &$ +0.630$ &$ -0.004$ &$ +2.538$ &$ +2.535$ &$ -0.003$ &$ -0.179$ &$ -0.179$ &$ +0.000$ &$ +1.008$ &$ +1.012$ &$ +0.004$\\
\end{tabular}
\end{center}
\end{table*}

\begin{table*}
\caption{\label{Gaiam00}
Colours and corrections for the Gaia system for metallicity [M/H]=0.0. In columns ATLAS$_c$ the 3D correction
has been added to the ATLAS colour. The $CaT$ colour is defined as $(G-RVS)-(BP-RP)$. $Bol_G$ is the bolometric
correction in the G-band. }
\renewcommand{\tabcolsep}{3pt}
\tabskip=0pt
\begin{center}
\begin{tabular}{llllllllllllll}
\hline\noalign{\smallskip}
\multicolumn{1}{c}{\teff} & 
\multicolumn{1}{c}{log g} & 
\multicolumn{1}{c}{$G-H_p$} &
\multicolumn{1}{c}{$G-H_p$} &
\multicolumn{1}{c}{$\Delta_{G-H_p}$} &
\multicolumn{1}{c}{$BP-RP$} &
\multicolumn{1}{c}{$BP-RP$} &
\multicolumn{1}{c}{$\Delta_{BP-RP}$} &
\multicolumn{1}{c}{$CaT$} &
\multicolumn{1}{c}{$CaT$} &
\multicolumn{1}{c}{$\Delta_{CaT}$} & 
\multicolumn{1}{c}{$Bol_G$} &
\multicolumn{1}{c}{$Bol_G$} &
\multicolumn{1}{c}{$\Delta_{Bol_G}$} \\
\multicolumn{2}{c}{ } &  
\multicolumn{1}{c}{ATLAS} &
\multicolumn{1}{c}{ATLAS$_c$}&
\multicolumn{1}{c}{3D-1D}&
\multicolumn{1}{c}{ATLAS} &
\multicolumn{1}{c}{ATLAS$_c$}&
\multicolumn{1}{c}{3D-1D}&
\multicolumn{1}{c}{ATLAS} &
\multicolumn{1}{c}{ATLAS$_c$}&
\multicolumn{1}{c}{3D-1D}&
\multicolumn{1}{c}{ATLAS} &
\multicolumn{1}{c}{ATLAS$_c$}&
\multicolumn{1}{c}{3D-1D} \\
\hline\noalign{\smallskip}
\hline\noalign{\smallskip}
3635  &1.00 & $-0.767$ &$ -0.780$ &$ -0.013$ &$ +2.247$ &$ +2.277$ &$ +0.030$ &$ -0.494$ &$ -0.511$  &$-0.017$ &$ -1.153$ &$ -1.272$&$  -0.119$ \\
3813  &4.00 & $-0.661$ &$ -0.655$ &$ +0.007$ &$ +2.022$ &$ +2.006$ &$ -0.016$ &$ -0.383$ &$ -0.377$  &$+0.006$ &$ -0.887$ &$ -0.857$&$  +0.029$ \\
4018  &1.50 & $-0.541$ &$ -0.531$ &$ +0.010$ &$ +1.752$ &$ +1.730$ &$ -0.021$ &$ -0.372$ &$ -0.359$  &$+0.013$ &$ -0.650$ &$ -0.628$&$  +0.022$ \\
3964  &4.50 & $-0.587$ &$ -0.586$ &$ +0.001$ &$ +1.852$ &$ +1.850$ &$ -0.001$ &$ -0.359$ &$ -0.349$  &$+0.010$ &$ -0.695$ &$ -0.687$&$  +0.008$ \\
4501  &2.00 & $-0.399$ &$ -0.394$ &$ +0.006$ &$ +1.415$ &$ +1.401$ &$ -0.014$ &$ -0.275$ &$ -0.269$  &$+0.006$ &$ -0.338$ &$ -0.322$&$  +0.016$ \\
4477  &2.50 & $-0.402$ &$ -0.400$ &$ +0.002$ &$ +1.422$ &$ +1.417$ &$ -0.005$ &$ -0.270$ &$ -0.267$  &$+0.003$ &$ -0.353$ &$ -0.346$&$  +0.007$ \\
\end{tabular}
\end{center}
\end{table*}

\begin{table*}
\caption{\label{Gaiam10}
Colours and corrections for the Gaia system for metallicity [M/H]=--1.0. In columns ATLAS$_c$ the 3D correction
has been added to the ATLAS colour. The $CaT$ colour is defined as $(G-RVS)-(BP-RP)$. $Bol_G$ is the bolometric
correction in the G-band. }
\renewcommand{\tabcolsep}{3pt}
\tabskip=0pt
\begin{center}
\begin{tabular}{llllllllllllll}
\hline\noalign{\smallskip}
\multicolumn{1}{c}{\teff} & 
\multicolumn{1}{c}{log g} & 
\multicolumn{1}{c}{$G-H_p$} &
\multicolumn{1}{c}{$G-H_p$} &
\multicolumn{1}{c}{$\Delta_{G-H_p}$} &
\multicolumn{1}{c}{$BP-RP$} &
\multicolumn{1}{c}{$BP-RP$} &
\multicolumn{1}{c}{$\Delta_{BP-RP}$} &
\multicolumn{1}{c}{$CaT$} &
\multicolumn{1}{c}{$CaT$} &
\multicolumn{1}{c}{$\Delta_{CaT}$} & 
\multicolumn{1}{c}{$Bol_G$} &
\multicolumn{1}{c}{$Bol_G$} &
\multicolumn{1}{c}{$\Delta_{Bol_G}$} \\
\multicolumn{2}{c}{ } &  
\multicolumn{1}{c}{ATLAS} &
\multicolumn{1}{c}{ATLAS$_c$}&
\multicolumn{1}{c}{3D-1D}&
\multicolumn{1}{c}{ATLAS} &
\multicolumn{1}{c}{ATLAS$_c$}&
\multicolumn{1}{c}{3D-1D}&
\multicolumn{1}{c}{ATLAS} &
\multicolumn{1}{c}{ATLAS$_c$}&
\multicolumn{1}{c}{3D-1D}&
\multicolumn{1}{c}{ATLAS} &
\multicolumn{1}{c}{ATLAS$_c$}&
\multicolumn{1}{c}{3D-1D} \\
\hline\noalign{\smallskip}
\hline\noalign{\smallskip}
3790  &4.00& $ -0.612$&$  -0.608$&$  +0.004$&$  +1.905$&$  +1.899$&$  -0.006$&$  -0.368$&$  -0.358$&$  +0.010$&$  -0.717$&$-0.705$&$+0.013$\\
4040  &1.50& $ -0.499$&$  -0.488$&$  +0.011$&$  +1.659$&$  +1.635$&$  -0.024$&$  -0.327$&$  -0.314$&$  +0.013$&$  -0.618$&$-0.570$&$+0.048$\\
4001  &4.50& $ -0.539$&$  -0.537$&$  +0.001$&$  +1.738$&$  +1.737$&$  -0.001$&$  -0.343$&$  -0.331$&$  +0.012$&$  -0.568$&$-0.566$&$+0.001$\\
4485  &2.00& $ -0.374$&$  -0.370$&$  +0.004$&$  +1.356$&$  +1.347$&$  -0.009$&$  -0.230$&$  -0.224$&$  +0.006$&$  -0.353$&$-0.342$&$+0.011$\\
4490  &2.50& $ -0.370$&$  -0.370$&$  -0.001$&$  +1.345$&$  +1.346$&$  +0.001$&$  -0.221$&$  -0.221$&$  +0.000$&$  -0.352$&$-0.346$&$+0.007$\\
4525  &4.00& $ -0.366$&$  -0.375$&$  -0.009$&$  +1.330$&$  +1.350$&$  +0.021$&$  -0.219$&$  -0.226$&$  -0.007$&$  -0.338$&$-0.344$&$-0.007$\\
\end{tabular}
\end{center}
\end{table*}

\begin{table*}
\caption{\label{Gaiam20}
Colours and corrections for the Gaia system for metallicity [M/H]=--2.0. In columns ATLAS$_c$ the 3D correction
has been added to the ATLAS colour. The $CaT$ colour is defined as $(G-RVS)-(BP-RP)$. $Bol_G$ is the bolometric
correction in the G-band. }
\renewcommand{\tabcolsep}{3pt}
\tabskip=0pt
\begin{center}
\begin{tabular}{llllllllllllll}
\hline\noalign{\smallskip}
\multicolumn{1}{c}{\teff} & 
\multicolumn{1}{c}{log g} & 
\multicolumn{1}{c}{$G-H_p$} &
\multicolumn{1}{c}{$G-H_p$} &
\multicolumn{1}{c}{$\Delta_{G-H_p}$} &
\multicolumn{1}{c}{$BP-RP$} &
\multicolumn{1}{c}{$BP-RP$} &
\multicolumn{1}{c}{$\Delta_{BP-RP}$} &
\multicolumn{1}{c}{$CaT$} &
\multicolumn{1}{c}{$CaT$} &
\multicolumn{1}{c}{$\Delta_{CaT}$} & 
\multicolumn{1}{c}{$Bol_G$} &
\multicolumn{1}{c}{$Bol_G$} &
\multicolumn{1}{c}{$\Delta_{Bol_G}$} \\
\multicolumn{2}{c}{ } &  
\multicolumn{1}{c}{ATLAS} &
\multicolumn{1}{c}{ATLAS$_c$}&
\multicolumn{1}{c}{3D-1D}&
\multicolumn{1}{c}{ATLAS} &
\multicolumn{1}{c}{ATLAS$_c$}&
\multicolumn{1}{c}{3D-1D}&
\multicolumn{1}{c}{ATLAS} &
\multicolumn{1}{c}{ATLAS$_c$}&
\multicolumn{1}{c}{3D-1D}&
\multicolumn{1}{c}{ATLAS} &
\multicolumn{1}{c}{ATLAS$_c$}&
\multicolumn{1}{c}{3D-1D} \\
\hline\noalign{\smallskip}
\hline\noalign{\smallskip}
4001  &1.50 & $-0.505$ &$ -0.501$ &$ +0.004$ &$ +1.677$ &$ +1.668$ &$ -0.009$ &$ -0.317$ &$ -0.314$ &$ +0.003$ &$ -0.646$&$ -0.600$&$ +0.046$\\
4000  &4.50 & $-0.515$ &$ -0.515$ &$ +0.000$ &$ +1.689$ &$ +1.689$ &$ +0.000$ &$ -0.327$ &$ -0.326$ &$ +0.001$ &$ -0.522$&$ -0.518$&$ +0.003$\\
4506  &1.50 & $-0.375$ &$ -0.370$ &$ +0.006$ &$ +1.361$ &$ +1.347$ &$ -0.014$ &$ -0.216$ &$ -0.207$ &$ +0.009$ &$ -0.352$&$ -0.348$&$ +0.005$\\
4506  &2.00 & $-0.362$ &$ -0.365$ &$ -0.003$ &$ +1.326$ &$ +1.333$ &$ +0.007$ &$ -0.196$ &$ -0.197$ &$ -0.001$ &$ -0.358$&$ -0.362$&$ -0.004$\\
4478  &2.50 & $-0.363$ &$ -0.368$ &$ -0.004$ &$ +1.330$ &$ +1.340$ &$ +0.010$ &$ -0.201$ &$ -0.206$ &$ -0.005$ &$ -0.349$&$ -0.346$&$ +0.002$\\
4504  &4.00 & $-0.360$ &$ -0.367$ &$ -0.007$ &$ +1.318$ &$ +1.334$ &$ +0.016$ &$ -0.198$ &$ -0.206$ &$ -0.008$ &$ -0.332$&$ -0.329$&$ +0.003$\\
\end{tabular}
\end{center}
\end{table*}

\begin{table*}
\caption{\label{Gaiam30}
Colours and corrections for the Gaia system for metallicity [M/H]=--3.0. In columns ATLAS$_c$ the 3D correction
has been added to the ATLAS colour. The $CaT$ colour is defined as $(G-RVS)-(BP-RP)$. $Bol_G$ is the bolometric
correction in the G-band. }
\renewcommand{\tabcolsep}{3pt}
\tabskip=0pt
\begin{center}
\begin{tabular}{llllllllllllll}
\hline\noalign{\smallskip}
\multicolumn{1}{c}{\teff} & 
\multicolumn{1}{c}{log g} & 
\multicolumn{1}{c}{$G-H_p$} &
\multicolumn{1}{c}{$G-H_p$} &
\multicolumn{1}{c}{$\Delta_{G-H_p}$} &
\multicolumn{1}{c}{$BP-RP$} &
\multicolumn{1}{c}{$BP-RP$} &
\multicolumn{1}{c}{$\Delta_{BP-RP}$} &
\multicolumn{1}{c}{$CaT$} &
\multicolumn{1}{c}{$CaT$} &
\multicolumn{1}{c}{$\Delta_{CaT}$} & 
\multicolumn{1}{c}{$Bol_G$} &
\multicolumn{1}{c}{$Bol_G$} &
\multicolumn{1}{c}{$\Delta_{Bol_G}$} \\
\multicolumn{2}{c}{ } &  
\multicolumn{1}{c}{ATLAS} &
\multicolumn{1}{c}{ATLAS$_c$}&
\multicolumn{1}{c}{3D-1D}&
\multicolumn{1}{c}{ATLAS} &
\multicolumn{1}{c}{ATLAS$_c$}&
\multicolumn{1}{c}{3D-1D}&
\multicolumn{1}{c}{ATLAS} &
\multicolumn{1}{c}{ATLAS$_c$}&
\multicolumn{1}{c}{3D-1D}&
\multicolumn{1}{c}{ATLAS} &
\multicolumn{1}{c}{ATLAS$_c$}&
\multicolumn{1}{c}{3D-1D} \\
\hline\noalign{\smallskip}
\hline\noalign{\smallskip}
3990  &1.50 & $-0.544$ &$ -0.544$ &$ +0.000$ &$ +1.766$ &$ +1.766$ &$ +0.000$ &$ -0.347$ &$ -0.347$ &$ +0.001$ &$ -0.661$& $-0.640$& $+0.021 $\\
4503  &1.50 & $-0.387$ &$ -0.379$ &$ +0.008$ &$ +1.388$ &$ +1.370$ &$ -0.018$ &$ -0.211$ &$ -0.200$ &$ +0.011$ &$ -0.361$& $-0.359$& $+0.003 $\\
4488  &2.00 & $-0.377$ &$ -0.379$ &$ -0.002$ &$ +1.364$ &$ +1.369$ &$ +0.005$ &$ -0.201$ &$ -0.201$ &$ +0.000$ &$ -0.358$& $-0.362$& $-0.005 $\\
4523  &2.50 & $-0.361$ &$ -0.363$ &$ -0.002$ &$ +1.322$ &$ +1.326$ &$ +0.005$ &$ -0.185$ &$ -0.187$ &$ -0.001$ &$ -0.336$& $-0.338$& $-0.003 $\\
4494  &4.00 & $-0.372$ &$ -0.375$ &$ -0.003$ &$ +1.348$ &$ +1.355$ &$ +0.006$ &$ -0.195$ &$ -0.198$ &$ -0.004$ &$ -0.317$& $-0.319$& $-0.002 $\\
4522  &4.50 & $-0.370$ &$ -0.371$ &$ -0.002$ &$ +1.342$ &$ +1.346$ &$ +0.004$ &$ -0.195$ &$ -0.197$ &$ -0.003$ &$ -0.300$& $-0.300$& $+0.001 $\\
\end{tabular}
\end{center}
\end{table*}

\clearpage

\begin{table*}
\caption{\label{SDSS_1_m00}
Colours and corrections for  SDSS $u-g$, $g-z$, $g-r$ and $g-i$ for metallicity [M/H]=0.0. In columns ATLAS$_c$ the 3D correction
has been added to the ATLAS colour.  }
\renewcommand{\tabcolsep}{3pt}
\tabskip=0pt
\begin{center}
\begin{tabular}{llllllllllllll}
\hline\noalign{\smallskip}
\multicolumn{1}{c}{\teff} & 
\multicolumn{1}{c}{log g} & 
\multicolumn{1}{c}{$u-g$} &
\multicolumn{1}{c}{$u-g$} &
\multicolumn{1}{c}{$\Delta_{u-g}$} &
\multicolumn{1}{c}{$g-z$} &
\multicolumn{1}{c}{$g-z$} &
\multicolumn{1}{c}{$\Delta_{g-z}$} &
\multicolumn{1}{c}{$g-r$} &
\multicolumn{1}{c}{$g-r$} &
\multicolumn{1}{c}{$\Delta_{g-r}$} & 
\multicolumn{1}{c}{$g-i$} &
\multicolumn{1}{c}{$g-i$} &
\multicolumn{1}{c}{$\Delta_{g-i}$} \\
\multicolumn{2}{c}{ } &  
\multicolumn{1}{c}{ATLAS} &
\multicolumn{1}{c}{ATLAS$_c$}&
\multicolumn{1}{c}{3D-1D}&
\multicolumn{1}{c}{ATLAS} &
\multicolumn{1}{c}{ATLAS$_c$}&
\multicolumn{1}{c}{3D-1D}&
\multicolumn{1}{c}{ATLAS} &
\multicolumn{1}{c}{ATLAS$_c$}&
\multicolumn{1}{c}{3D-1D}&
\multicolumn{1}{c}{ATLAS} &
\multicolumn{1}{c}{ATLAS$_c$}&
\multicolumn{1}{c}{3D-1D} \\
\hline\noalign{\smallskip}
\hline\noalign{\smallskip}
3635  &1.00 & $+3.578$ & $+3.479$ & $-0.099$ & $+2.869$ & $+2.906$ & $+0.037$ & $+1.492$ & $+1.485$ & $-0.007$ & $+2.340$ & $+2.355$ & $+0.014$ \\
3813  &4.00 & $+2.579$ & $+2.542$ & $-0.037$ & $+2.484$ & $+2.468$ & $-0.016$ & $+1.294$ & $+1.312$ & $+0.017$ & $+2.031$ & $+2.025$ & $-0.007$ \\
4018  &1.50 & $+3.324$ & $+3.197$ & $-0.127$ & $+2.075$ & $+2.030$ & $-0.045$ & $+1.322$ & $+1.280$ & $-0.042$ & $+1.781$ & $+1.737$ & $-0.044$ \\
3964  &4.50 & $+2.548$ & $+2.483$ & $-0.066$ & $+2.233$ & $+2.226$ & $-0.008$ & $+1.301$ & $+1.275$ & $-0.025$ & $+1.879$ & $+1.864$ & $-0.015$ \\
4501  &2.00 & $+2.575$ & $+2.502$ & $-0.073$ & $+1.470$ & $+1.445$ & $-0.025$ & $+0.998$ & $+0.978$ & $-0.020$ & $+1.297$ & $+1.275$ & $-0.022$ \\
4477  &2.50 & $+2.487$ & $+2.457$ & $-0.030$ & $+1.485$ & $+1.475$ & $-0.010$ & $+0.997$ & $+0.987$ & $-0.010$ & $+1.306$ & $+1.297$ & $-0.009$ \\
\end{tabular}
\end{center}
\end{table*}

\begin{table*}
\caption{\label{SDSS_1_m10}
Colours and corrections for  SDSS $u-g$, $g-z$, $g-r$ and $g-i$ for metallicity [M/H]=--1.0. In columns ATLAS$_c$ the 3D correction
has been added to the ATLAS colour.  }
\renewcommand{\tabcolsep}{3pt}
\tabskip=0pt
\begin{center}
\begin{tabular}{llllllllllllll}
\hline\noalign{\smallskip}
\multicolumn{1}{c}{\teff} & 
\multicolumn{1}{c}{log g} & 
\multicolumn{1}{c}{$u-g$} &
\multicolumn{1}{c}{$u-g$} &
\multicolumn{1}{c}{$\Delta_{u-g}$} &
\multicolumn{1}{c}{$g-z$} &
\multicolumn{1}{c}{$g-z$} &
\multicolumn{1}{c}{$\Delta_{g-z}$} &
\multicolumn{1}{c}{$g-r$} &
\multicolumn{1}{c}{$g-r$} &
\multicolumn{1}{c}{$\Delta_{g-r}$} & 
\multicolumn{1}{c}{$g-i$} &
\multicolumn{1}{c}{$g-i$} &
\multicolumn{1}{c}{$\Delta_{g-i}$} \\
\multicolumn{2}{c}{ } &  
\multicolumn{1}{c}{ATLAS} &
\multicolumn{1}{c}{ATLAS$_c$}&
\multicolumn{1}{c}{3D-1D}&
\multicolumn{1}{c}{ATLAS} &
\multicolumn{1}{c}{ATLAS$_c$}&
\multicolumn{1}{c}{3D-1D}&
\multicolumn{1}{c}{ATLAS} &
\multicolumn{1}{c}{ATLAS$_c$}&
\multicolumn{1}{c}{3D-1D}&
\multicolumn{1}{c}{ATLAS} &
\multicolumn{1}{c}{ATLAS$_c$}&
\multicolumn{1}{c}{3D-1D} \\
\hline\noalign{\smallskip}
\hline\noalign{\smallskip}
3790  &4.00 & $+2.311$ & $+2.272$ & $-0.039$ & $+2.334$ & $+2.318$ & $-0.016$ & $+1.346$ & $+1.320$ & $-0.026$ & $+1.951$ & $+1.930$ & $-0.021$ \\
4040  &1.50 & $+2.575$ & $+2.488$ & $-0.086$ & $+1.890$ & $+1.845$ & $-0.046$ & $+1.171$ & $+1.134$ & $-0.037$ & $+1.615$ & $+1.573$ & $-0.042$ \\
4001  &4.50 & $+2.208$ & $+2.169$ & $-0.039$ & $+2.052$ & $+2.046$ & $-0.006$ & $+1.270$ & $+1.242$ & $-0.027$ & $+1.758$ & $+1.741$ & $-0.017$ \\
4485  &2.00 & $+1.938$ & $+1.903$ & $-0.035$ & $+1.370$ & $+1.353$ & $-0.016$ & $+0.865$ & $+0.851$ & $-0.013$ & $+1.186$ & $+1.171$ & $-0.014$ \\
4490  &2.50 & $+1.847$ & $+1.841$ & $-0.006$ & $+1.355$ & $+1.357$ & $+0.002$ & $+0.853$ & $+0.853$ & $+0.000$ & $+1.171$ & $+1.174$ & $+0.002$ \\
4525  &4.00 & $+1.735$ & $+1.760$ & $+0.025$ & $+1.344$ & $+1.380$ & $+0.036$ & $+0.860$ & $+0.884$ & $+0.024$ & $+1.168$ & $+1.201$ & $+0.032$ \\
\end{tabular}
\end{center}
\end{table*}

\begin{table*}
\caption{\label{SDSS_1_m20}
Colours and corrections for  SDSS $u-g$, $g-z$, $g-r$ and $g-i$ for metallicity [M/H]=--2.0. In columns ATLAS$_c$ the 3D correction
has been added to the ATLAS colour.  }
\renewcommand{\tabcolsep}{3pt}
\tabskip=0pt
\begin{center}
\begin{tabular}{llllllllllllll}
\hline\noalign{\smallskip}
\multicolumn{1}{c}{\teff} & 
\multicolumn{1}{c}{log g} & 
\multicolumn{1}{c}{$u-g$} &
\multicolumn{1}{c}{$u-g$} &
\multicolumn{1}{c}{$\Delta_{u-g}$} &
\multicolumn{1}{c}{$g-z$} &
\multicolumn{1}{c}{$g-z$} &
\multicolumn{1}{c}{$\Delta_{g-z}$} &
\multicolumn{1}{c}{$g-r$} &
\multicolumn{1}{c}{$g-r$} &
\multicolumn{1}{c}{$\Delta_{g-r}$} & 
\multicolumn{1}{c}{$g-i$} &
\multicolumn{1}{c}{$g-i$} &
\multicolumn{1}{c}{$\Delta_{g-i}$} \\
\multicolumn{2}{c}{ } &  
\multicolumn{1}{c}{ATLAS} &
\multicolumn{1}{c}{ATLAS$_c$}&
\multicolumn{1}{c}{3D-1D}&
\multicolumn{1}{c}{ATLAS} &
\multicolumn{1}{c}{ATLAS$_c$}&
\multicolumn{1}{c}{3D-1D}&
\multicolumn{1}{c}{ATLAS} &
\multicolumn{1}{c}{ATLAS$_c$}&
\multicolumn{1}{c}{3D-1D}&
\multicolumn{1}{c}{ATLAS} &
\multicolumn{1}{c}{ATLAS$_c$}&
\multicolumn{1}{c}{3D-1D} \\
\hline\noalign{\smallskip}
\hline\noalign{\smallskip}
4001  &1.50 & $+2.411$ & $+2.364$ & $-0.047$ & $+1.912$ & $+1.894$ & $-0.019$ & $+1.158$ & $+1.148$ & $-0.010$ & $+1.623 $& $+1.610$ & $-0.013$ \\
4000  &4.50 & $+2.076$ & $+2.072$ & $-0.003$ & $+1.951$ & $+1.950$ & $-0.001$ & $+1.209$ & $+1.207$ & $-0.002$ & $+1.676 $& $+1.675$ & $-0.001$ \\
4506  &1.50 & $+1.824$ & $+1.784$ & $-0.040$ & $+1.375$ & $+1.351$ & $-0.024$ & $+0.851$ & $+0.828$ & $-0.024$ & $+1.185 $& $+1.162$ & $-0.023$ \\
4506  &2.00 & $+1.644$ & $+1.639$ & $-0.006$ & $+1.327$ & $+1.338$ & $+0.011$ & $+0.810$ & $+0.814$ & $+0.004$ & $+1.135 $& $+1.145$ & $+0.010$ \\
4478  &2.50 & $+1.701$ & $+1.706$ & $+0.005$ & $+1.330$ & $+1.346$ & $+0.016$ & $+0.817$ & $+0.830$ & $+0.013$ & $+1.142 $& $+1.159$ & $+0.017$ \\
4504  &4.00 & $+1.589$ & $+1.605$ & $+0.016$ & $+1.320$ & $+1.345$ & $+0.025$ & $+0.818$ & $+0.839$ & $+0.021$ & $+1.137 $& $+1.163$ & $+0.026$ \\
\end{tabular}
\end{center}
\end{table*}

\begin{table*}
\caption{\label{SDSS_1_m30}
Colours and corrections for  SDSS $u-g$, $g-z$, $g-r$ and $g-i$ for metallicity [M/H]=--3.0. In columns ATLAS$_c$ the 3D correction
has been added to the ATLAS colour.  }
\renewcommand{\tabcolsep}{3pt}
\tabskip=0pt
\begin{center}
\begin{tabular}{llllllllllllll}
\hline\noalign{\smallskip}
\multicolumn{1}{c}{\teff} & 
\multicolumn{1}{c}{log g} & 
\multicolumn{1}{c}{$u-g$} &
\multicolumn{1}{c}{$u-g$} &
\multicolumn{1}{c}{$\Delta_{u-g}$} &
\multicolumn{1}{c}{$g-z$} &
\multicolumn{1}{c}{$g-z$} &
\multicolumn{1}{c}{$\Delta_{g-z}$} &
\multicolumn{1}{c}{$g-r$} &
\multicolumn{1}{c}{$g-r$} &
\multicolumn{1}{c}{$\Delta_{g-r}$} & 
\multicolumn{1}{c}{$g-i$} &
\multicolumn{1}{c}{$g-i$} &
\multicolumn{1}{c}{$\Delta_{g-i}$} \\
\multicolumn{2}{c}{ } &  
\multicolumn{1}{c}{ATLAS} &
\multicolumn{1}{c}{ATLAS$_c$}&
\multicolumn{1}{c}{3D-1D}&
\multicolumn{1}{c}{ATLAS} &
\multicolumn{1}{c}{ATLAS$_c$}&
\multicolumn{1}{c}{3D-1D}&
\multicolumn{1}{c}{ATLAS} &
\multicolumn{1}{c}{ATLAS$_c$}&
\multicolumn{1}{c}{3D-1D}&
\multicolumn{1}{c}{ATLAS} &
\multicolumn{1}{c}{ATLAS$_c$}&
\multicolumn{1}{c}{3D-1D} \\
\hline\noalign{\smallskip}
\hline\noalign{\smallskip}
3990  &1.50 & $+2.543$ & $+2.560$ & $+0.018$ & $+2.054$ & $+2.050$ & $-0.004$ & $+1.241$ & $+1.237$ & $-0.004$ & $+1.748$ & $+1.746$ & $-0.002$ \\
4503  &1.50 & $+1.748$ & $+1.691$ & $-0.058$ & $+1.421$ & $+1.389$ & $-0.032$ & $+0.868$ & $+0.837$ & $-0.031$ & $+1.220$ & $+1.189$ & $-0.032$ \\
4488  &2.00 & $+1.660$ & $+1.647$ & $-0.013$ & $+1.385$ & $+1.391$ & $+0.006$ & $+0.841$ & $+0.841$ & $-0.001$ & $+1.186$ & $+1.191$ & $+0.005$ \\
4523  &2.50 & $+1.544$ & $+1.530$ & $-0.014$ & $+1.320$ & $+1.326$ & $+0.006$ & $+0.800$ & $+0.803$ & $+0.003$ & $+1.130$ & $+1.136$ & $+0.007$ \\
4494  &4.00 & $+1.516$ & $+1.526$ & $+0.009$ & $+1.361$ & $+1.371$ & $+0.010$ & $+0.826$ & $+0.835$ & $+0.009$ & $+1.169$ & $+1.180$ & $+0.011$ \\
4522  &4.50 & $+1.506$ & $+1.513$ & $+0.008$ & $+1.349$ & $+1.355$ & $+0.006$ & $+0.823$ & $+0.829$ & $+0.006$ & $+1.163$ & $+1.170$ & $+0.007$ \\
\end{tabular}
\end{center}
\end{table*}

\begin{table*}
\caption{\label{SDSS_2_m00}
Colours and corrections for  SDSS $r-i$, $i-z$, $u-i$ and $u-z$ for metallicity [M/H]=0.0. In columns ATLAS$_c$ the 3D correction
has been added to the ATLAS colour.  }
\renewcommand{\tabcolsep}{3pt}
\tabskip=0pt
\begin{center}
\begin{tabular}{llllllllllllll}
\hline\noalign{\smallskip}
\multicolumn{1}{c}{\teff} & 
\multicolumn{1}{c}{log g} & 
\multicolumn{1}{c}{$r-i$} &
\multicolumn{1}{c}{$r-i$} &
\multicolumn{1}{c}{$\Delta_{r-i}$} &
\multicolumn{1}{c}{$i-z$} &
\multicolumn{1}{c}{$i-z$} &
\multicolumn{1}{c}{$\Delta_{i-z}$} &
\multicolumn{1}{c}{$u-i$} &
\multicolumn{1}{c}{$u-i$} &
\multicolumn{1}{c}{$\Delta_{u-i}$} & 
\multicolumn{1}{c}{$u-z$} &
\multicolumn{1}{c}{$u-z$} &
\multicolumn{1}{c}{$\Delta_{u-z}$} \\
\multicolumn{2}{c}{ } &  
\multicolumn{1}{c}{ATLAS} &
\multicolumn{1}{c}{ATLAS$_c$}&
\multicolumn{1}{c}{3D-1D}&
\multicolumn{1}{c}{ATLAS} &
\multicolumn{1}{c}{ATLAS$_c$}&
\multicolumn{1}{c}{3D-1D}&
\multicolumn{1}{c}{ATLAS} &
\multicolumn{1}{c}{ATLAS$_c$}&
\multicolumn{1}{c}{3D-1D}&
\multicolumn{1}{c}{ATLAS} &
\multicolumn{1}{c}{ATLAS$_c$}&
\multicolumn{1}{c}{3D-1D} \\
\hline\noalign{\smallskip}
\hline\noalign{\smallskip}
3635  &1.00 & $+0.848$ & $+0.869$ & $+0.021$ & $+0.529$ & $+0.551$ & $+0.023$ & $+5.918$ & $+5.833$ & $-0.084$ & $+6.446$ & $+6.385$ & $-0.062$ \\
3813  &4.00 & $+0.737$ & $+0.713$ & $-0.024$ & $+0.452$ & $+0.443$ & $-0.009$ & $+4.611$ & $+4.567$ & $-0.044$ & $+5.063$ & $+5.010$ & $-0.053$ \\
4018  &1.50 & $+0.459$ & $+0.457$ & $-0.002$ & $+0.293$ & $+0.292$ & $-0.001$ & $+5.105$ & $+4.935$ & $-0.171$ & $+5.399$ & $+5.227$ & $-0.172$ \\
3964  &4.50 & $+0.578$ & $+0.588$ & $+0.010$ & $+0.354$ & $+0.362$ & $+0.008$ & $+4.427$ & $+4.346$ & $-0.081$ & $+4.782$ & $+4.708$ & $-0.073$ \\
4501  &2.00 & $+0.299$ & $+0.298$ & $-0.002$ & $+0.173$ & $+0.170$ & $-0.003$ & $+3.872$ & $+3.778$ & $-0.094$ & $+4.045$ & $+3.948$ & $-0.098$ \\
4477  &2.50 & $+0.309$ & $+0.309$ & $+0.001$ & $+0.179$ & $+0.178$ & $-0.001$ & $+3.793$ & $+3.754$ & $-0.039$ & $+3.972$ & $+3.932$ & $-0.040$ \\
\end{tabular}
\end{center}
\end{table*}

\begin{table*}
\caption{\label{SDSS_2_m10}
Colours and corrections for  SDSS $r-i$, $i-z$, $u-i$ and $u-z$ for metallicity [M/H]=--1.0. In columns ATLAS$_c$ the 3D correction
has been added to the ATLAS colour.  }
\renewcommand{\tabcolsep}{3pt}
\tabskip=0pt
\begin{center}
\begin{tabular}{llllllllllllll}
\hline\noalign{\smallskip}
\multicolumn{1}{c}{\teff} & 
\multicolumn{1}{c}{log g} & 
\multicolumn{1}{c}{$r-i$} &
\multicolumn{1}{c}{$r-i$} &
\multicolumn{1}{c}{$\Delta_{r-i}$} &
\multicolumn{1}{c}{$i-z$} &
\multicolumn{1}{c}{$i-z$} &
\multicolumn{1}{c}{$\Delta_{i-z}$} &
\multicolumn{1}{c}{$u-i$} &
\multicolumn{1}{c}{$u-i$} &
\multicolumn{1}{c}{$\Delta_{u-i}$} & 
\multicolumn{1}{c}{$u-z$} &
\multicolumn{1}{c}{$u-z$} &
\multicolumn{1}{c}{$\Delta_{u-z}$} \\
\multicolumn{2}{c}{ } &  
\multicolumn{1}{c}{ATLAS} &
\multicolumn{1}{c}{ATLAS$_c$}&
\multicolumn{1}{c}{3D-1D}&
\multicolumn{1}{c}{ATLAS} &
\multicolumn{1}{c}{ATLAS$_c$}&
\multicolumn{1}{c}{3D-1D}&
\multicolumn{1}{c}{ATLAS} &
\multicolumn{1}{c}{ATLAS$_c$}&
\multicolumn{1}{c}{3D-1D}&
\multicolumn{1}{c}{ATLAS} &
\multicolumn{1}{c}{ATLAS$_c$}&
\multicolumn{1}{c}{3D-1D} \\
\hline\noalign{\smallskip}
\hline\noalign{\smallskip}
3790  &4.00 & $+0.605$ & $+0.610$ & $+0.005$ & $+0.383$ & $+0.388$ & $+0.005$ & $+4.262$ & $+4.201$ & $-0.060$ & $+4.645$ & $+4.589$ & $-0.056$ \\
4040  &1.50 & $+0.444$ & $+0.439$ & $-0.005$ & $+0.276$ & $+0.272$ & $-0.004$ & $+4.189$ & $+4.061$ & $-0.128$ & $+4.465$ & $+4.333$ & $-0.132$ \\
4001  &4.50 & $+0.488$ & $+0.499$ & $+0.011$ & $+0.294$ & $+0.305$ & $+0.011$ & $+3.966$ & $+3.911$ & $-0.055$ & $+4.260$ & $+4.216$ & $-0.044$ \\
4485  &2.00 & $+0.321$ & $+0.320$ & $-0.001$ & $+0.184$ & $+0.182$ & $-0.002$ & $+3.124$ & $+3.074$ & $-0.050$ & $+3.308$ & $+3.256$ & $-0.052$ \\
4490  &2.50 & $+0.318$ & $+0.320$ & $+0.002$ & $+0.184$ & $+0.183$ & $-0.001$ & $+3.018$ & $+3.014$ & $-0.004$ & $+3.202$ & $+3.198$ & $-0.005$ \\
4525  &4.00 & $+0.308$ & $+0.316$ & $+0.008$ & $+0.176$ & $+0.180$ & $+0.003$ & $+2.903$ & $+2.960$ & $+0.058$ & $+3.079$ & $+3.140$ & $+0.061$ \\
\end{tabular}
\end{center}
\end{table*}

\begin{table*}
\caption{\label{SDSS_2_m20}
Colours and corrections for  SDSS $r-i$, $i-z$, $u-i$ and $u-z$ for metallicity [M/H]=--2.0. In columns ATLAS$_c$ the 3D correction
has been added to the ATLAS colour.  }
\renewcommand{\tabcolsep}{3pt}
\tabskip=0pt
\begin{center}
\begin{tabular}{llllllllllllll}
\hline\noalign{\smallskip}
\multicolumn{1}{c}{\teff} & 
\multicolumn{1}{c}{log g} & 
\multicolumn{1}{c}{$r-i$} &
\multicolumn{1}{c}{$r-i$} &
\multicolumn{1}{c}{$\Delta_{r-i}$} &
\multicolumn{1}{c}{$i-z$} &
\multicolumn{1}{c}{$i-z$} &
\multicolumn{1}{c}{$\Delta_{i-z}$} &
\multicolumn{1}{c}{$u-i$} &
\multicolumn{1}{c}{$u-i$} &
\multicolumn{1}{c}{$\Delta_{u-i}$} & 
\multicolumn{1}{c}{$u-z$} &
\multicolumn{1}{c}{$u-z$} &
\multicolumn{1}{c}{$\Delta_{u-z}$} \\
\multicolumn{2}{c}{ } &  
\multicolumn{1}{c}{ATLAS} &
\multicolumn{1}{c}{ATLAS$_c$}&
\multicolumn{1}{c}{3D-1D}&
\multicolumn{1}{c}{ATLAS} &
\multicolumn{1}{c}{ATLAS$_c$}&
\multicolumn{1}{c}{3D-1D}&
\multicolumn{1}{c}{ATLAS} &
\multicolumn{1}{c}{ATLAS$_c$}&
\multicolumn{1}{c}{3D-1D}&
\multicolumn{1}{c}{ATLAS} &
\multicolumn{1}{c}{ATLAS$_c$}&
\multicolumn{1}{c}{3D-1D} \\
\hline\noalign{\smallskip}
\hline\noalign{\smallskip}
4001  &1.50 & $+0.464$ & $+0.462$ & $-0.002$ & $+0.290$ & $+0.284$ & $-0.006$ & $+4.033$ & $+3.974$ & $-0.060$ & $+4.323$ & $+4.257$ & $-0.066$ \\
4000  &4.50 & $+0.467$ & $+0.468$ & $+0.001$ & $+0.275$ & $+0.275$ & $+0.000$ & $+3.752$ & $+3.747$ & $-0.004$ & $+4.026$ & $+4.022$ & $-0.004$ \\
4506  &1.50 & $+0.334$ & $+0.334$ & $+0.000$ & $+0.190$ & $+0.190$ & $-0.001$ & $+3.009$ & $+2.945$ & $-0.063$ & $+3.199$ & $+3.135$ & $-0.064$ \\
4506  &2.00 & $+0.325$ & $+0.331$ & $+0.006$ & $+0.192$ & $+0.192$ & $+0.001$ & $+2.780$ & $+2.784$ & $+0.004$ & $+2.971$ & $+2.976$ & $+0.005$ \\
4478  &2.50 & $+0.325$ & $+0.329$ & $+0.004$ & $+0.188$ & $+0.187$ & $-0.001$ & $+2.843$ & $+2.865$ & $+0.022$ & $+3.031$ & $+3.052$ & $+0.020$ \\
4504  &4.00 & $+0.319$ & $+0.324$ & $+0.005$ & $+0.183$ & $+0.182$ & $-0.001$ & $+2.726$ & $+2.768$ & $+0.041$ & $+2.910$ & $+2.950$ & $+0.040$ \\
\end{tabular}
\end{center}
\end{table*}

\begin{table*}
\caption{\label{SDSS_2_m30}
Colours and corrections for  SDSS $r-i$, $i-z$, $u-i$ and $u-z$ for metallicity [M/H]=--3.0. In columns ATLAS$_c$ the 3D correction
has been added to the ATLAS colour.  }
\renewcommand{\tabcolsep}{3pt}
\tabskip=0pt
\begin{center}
\begin{tabular}{llllllllllllll}
\hline\noalign{\smallskip}
\multicolumn{1}{c}{\teff} & 
\multicolumn{1}{c}{log g} & 
\multicolumn{1}{c}{$r-i$} &
\multicolumn{1}{c}{$r-i$} &
\multicolumn{1}{c}{$\Delta_{r-i}$} &
\multicolumn{1}{c}{$i-z$} &
\multicolumn{1}{c}{$i-z$} &
\multicolumn{1}{c}{$\Delta_{i-z}$} &
\multicolumn{1}{c}{$u-i$} &
\multicolumn{1}{c}{$u-i$} &
\multicolumn{1}{c}{$\Delta_{u-i}$} & 
\multicolumn{1}{c}{$u-z$} &
\multicolumn{1}{c}{$u-z$} &
\multicolumn{1}{c}{$\Delta_{u-z}$} \\
\multicolumn{2}{c}{ } &  
\multicolumn{1}{c}{ATLAS} &
\multicolumn{1}{c}{ATLAS$_c$}&
\multicolumn{1}{c}{3D-1D}&
\multicolumn{1}{c}{ATLAS} &
\multicolumn{1}{c}{ATLAS$_c$}&
\multicolumn{1}{c}{3D-1D}&
\multicolumn{1}{c}{ATLAS} &
\multicolumn{1}{c}{ATLAS$_c$}&
\multicolumn{1}{c}{3D-1D}&
\multicolumn{1}{c}{ATLAS} &
\multicolumn{1}{c}{ATLAS$_c$}&
\multicolumn{1}{c}{3D-1D} \\
\hline\noalign{\smallskip}
\hline\noalign{\smallskip}
3990  &1.50 & $+0.507$ & $+0.509$ & $+0.002$ & $+0.306$ & $+0.304$ & $-0.002$ & $+4.290$ & $+4.306$ & $+0.016$ & $+4.596$ & $+4.611$ & $+0.014$ \\
4503  &1.50 & $+0.352$ & $+0.352$ & $-0.000$ & $+0.201$ & $+0.200$ & $-0.000$ & $+2.969$ & $+2.879$ & $-0.089$ & $+3.169$ & $+3.080$ & $-0.090$ \\
4488  &2.00 & $+0.344$ & $+0.350$ & $+0.006$ & $+0.199$ & $+0.200$ & $+0.001$ & $+2.846$ & $+2.838$ & $-0.008$ & $+3.045$ & $+3.038$ & $-0.007$ \\
4523  &2.50 & $+0.329$ & $+0.334$ & $+0.004$ & $+0.190$ & $+0.190$ & $-0.000$ & $+2.674$ & $+2.666$ & $-0.007$ & $+2.864$ & $+2.856$ & $-0.008$ \\
4494  &4.00 & $+0.343$ & $+0.345$ & $+0.002$ & $+0.192$ & $+0.191$ & $-0.000$ & $+2.686$ & $+2.706$ & $+0.020$ & $+2.877$ & $+2.897$ & $+0.020$ \\
4522  &4.50 & $+0.340$ & $+0.340$ & $+0.001$ & $+0.186$ & $+0.186$ & $-0.001$ & $+2.668$ & $+2.683$ & $+0.015$ & $+2.855$ & $+2.869$ & $+0.014$ \\
\end{tabular}
\end{center}
\end{table*}

\begin{figure*}[t!]
\resizebox{0.5\hsize}{!}{\includegraphics[clip=true]{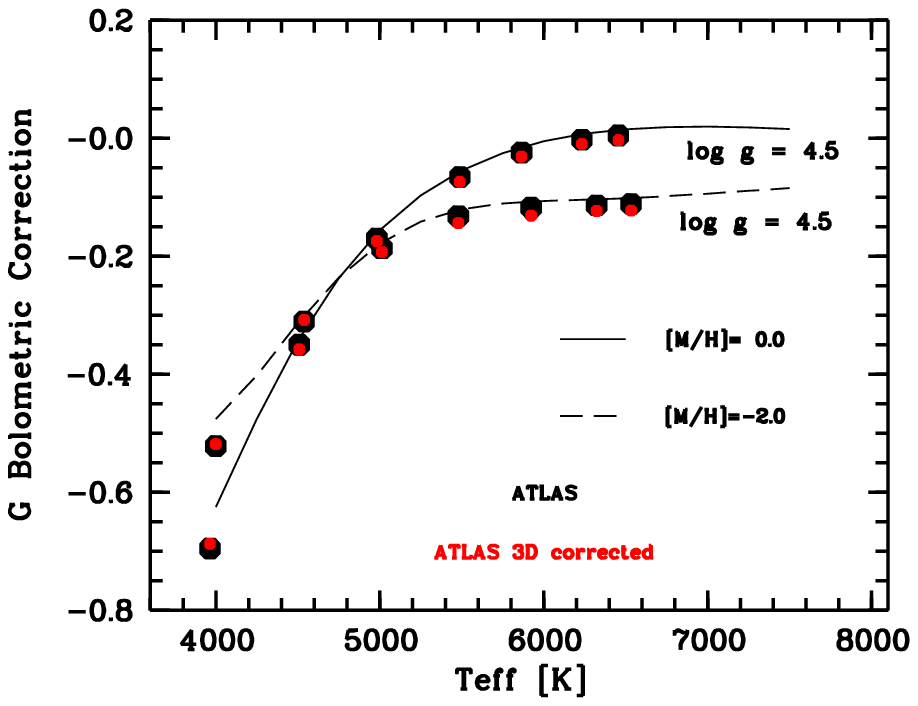}}
\resizebox{0.5\hsize}{!}{\includegraphics[clip=true]{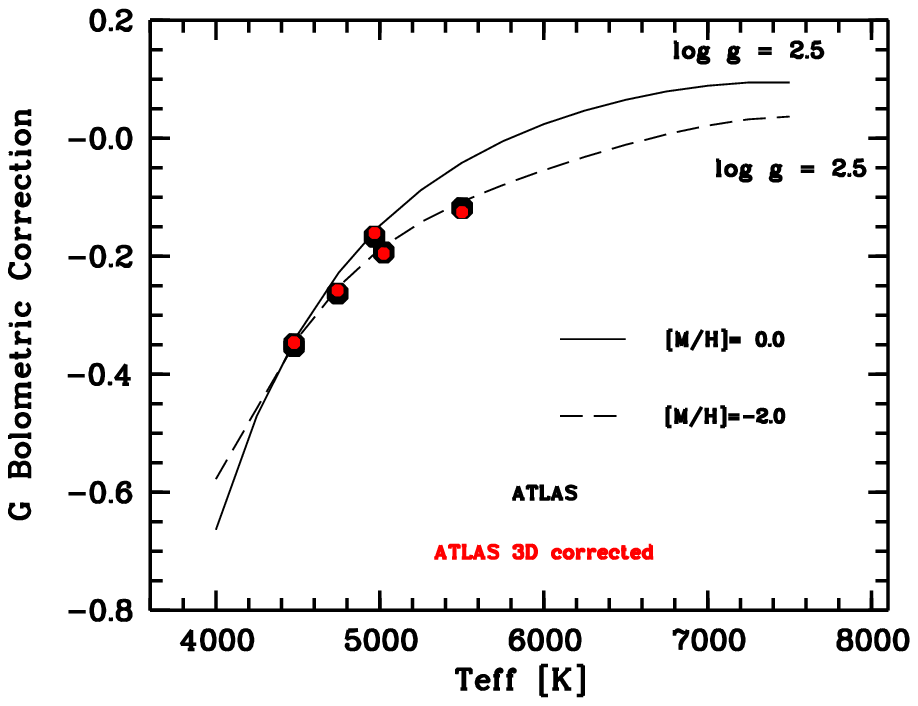}}
\caption{\footnotesize
Bolometric corrections in the Gaia G band for four sets of models:
log g = 2.5 (right panel) and log = 4.5 (left panel) at [M/H]=0.0 and [M/H]=--2.0.
Black symbols are computed from our ATLAS models, the lines
are the bolometric corrections computed from the \citet{CK03} grid.
The solid line refers to models of solar metallicity, the dashed
line to models with metallicity --2.0. The red symbols are
the values to which we added the 3D correction.
}
\label{BolG}
\end{figure*}

We show in Fig.\ref{BolG}
the bolometric corrections to Gaia $G$ magnitude as a function
of effective temperature. 
This Figure supersedes and replaces the analogous Figure 2 of \citet{Bonifacio_Mem}, in which
a large correction was predicted for the hottest solar metallicity models. Those
corrections were in fact spurious and due to the use of  low  resolution LHD models.
It can be appreciated as for solar-type solar-metallicity stars most
of the stellar flux is emitted in the $G$ bandpass, leading to essentially zero
bolometric corrections. For cooler stars a larger part of the flux
is emitted in the near infra-red, outside the $G$ bandpass and thus
the bolometric correction becomes non-neglegible.
In a similar manner we note that for metal-poor stars, even of solar-type,
the bolometric correction is small, but non-neglegible. This is due to the fact
that because of the lower opacity more flux is emitted in the 
ultra-violet, outside the $G$ bandpass.

\end{appendix}

\end{document}